\documentclass[aps,prx,twocolumn,superscriptaddress,amsmath,amssymb]{revtex4-1}
\usepackage{graphicx}
\usepackage{float}
\usepackage{bm} % for math
\usepackage{verbatim}  % for math
\usepackage{mathrsfs}
\usepackage{xcolor}

\newcommand{\alex}{\color{black}}

\begin{document}

\title{Swimming suppresses correlations in dilute suspensions of pusher microorganisms}

\author{Viktor \v{S}kult\'ety}
\affiliation{SUPA, School of Physics and Astronomy, The University of Edinburgh, James Clerk Maxwell Building, Peter Guthrie Tait Road, Edinburgh, EH9 3FD, United Kingdom}

\author{Cesare Nardini}
\affiliation{Service de Physique de l'\'{E}tat Condens\'{e}, CNRS UMR 3680, CEA-Saclay, 91191 Gif-sur-Yvette, France}

\author{Joakim Stenhammar}
\affiliation{Division of Physical Chemistry, Lund University, Box 124, S-221 00 Lund, Sweden}

\author{Davide Marenduzzo}
\affiliation{SUPA, School of Physics and Astronomy, The University of Edinburgh, James Clerk Maxwell Building, Peter Guthrie Tait Road, Edinburgh, EH9 3FD, United Kingdom}

\author{Alexander Morozov}
\email{alexander.morozov@ed.ac.uk}
\affiliation{SUPA, School of Physics and Astronomy, The University of Edinburgh, James Clerk Maxwell Building, Peter Guthrie Tait Road, Edinburgh, EH9 3FD, United Kingdom}

\date{\today}
\begin{abstract}
Active matter exhibits various forms of non-equilibrium states in the absence of external forcing, including macroscopic steady-state currents. Such states are often too complex to be modelled from first principles and our understanding of their physics relies heavily on minimal models. These have mostly been studied in the case of ``dry" active matter, where particle dynamics are dominated by friction with their surroundings.  Significantly less is known about systems with long-range hydrodynamic interactions that belong to ``wet" active matter. Dilute suspensions of motile bacteria, modelled as self-propelled dipolar particles interacting solely through long-ranged hydrodynamic fields, are arguably the most studied example from this class of active systems. Their phenomenology is well-established: at sufficiently high density of bacteria, there appear large-scale vortices and jets comprising many individual organisms, forming a chaotic state commonly known as \emph{bacterial turbulence}. As revealed by computer simulations, below the onset of collective motion, the suspension exhibits very strong correlations between individual microswimmers stemming from the long-ranged nature of dipolar fields. Here we demonstrate that this phenomenology is captured by the minimal model of microswimmers.  We develop a kinetic theory that goes beyond the commonly used mean-field assumption, and explicitly takes into account such correlations. Notably, these can be computed exactly within our theory. We calculate the fluid velocity variance, spatial and temporal correlation functions, the fluid velocity spectrum, and the enhanced diffusivity of tracer particles. 
We find that correlations are suppressed by particle self-propulsion, although the mean-field behaviour is not restored even in the limit of very fast swimming. Our theory is not perturbative and is valid for any value of the micro-swimmer density below the onset of collective motion. This work constitutes a significant methodological advance and allows us to make qualitative and quantitative predictions that can be directly compared to experiments and computer simulations of micro-swimmer suspensions.
\end{abstract}

\maketitle

\section{Introduction}
\label{section:introduction}

In recent years active systems emerged as a new state of matter with unique properties that are absent from their passive counterparts \cite{Ramaswamy2010,Marchetti2013}. Such systems comprise particles that are capable of extracting energy from their environment and using it to exert forces and torques on their surroundings. The resulting self-propulsion and interactions between particles break detailed balance at the microscopic level, often leading to steady states that are not invariant under time reversal and exhibit macroscopic currents \cite{Cates2012}. Such currents, or collective motion, have been reported in a variety of systems \cite{Vicsek2012}, including Vicsek particles \cite{Chate2008}, mixtures of microtubules and molecular motors \cite{Sanchez2012}, light-activated colloids \cite{Palacci2013}, Quincke rollers \cite{Bricard2013,Karani2019}, bacterial colonies \cite{Zhang2010}, sperm cells \cite{Creppy2015}, locusts \cite{Buhl2006}, birds, and fish \cite{Parrish1997}. The omnipresence of collective motion raises the need to classify various active systems according to common features of their phenomenological behaviour. Marchetti \emph{et al.} \cite{Marchetti2013} recently introduced two broad universality classes for active systems, ``dry" and ``wet", comprising particles dominated by friction with their surroundings and long-ranged hydrodynamic interactions, respectively. Each class is expected to be defined by a few, relatively simple model systems, and significant effort has been invested into finding such models. 
For dry active matter, these include Vicsek-like models \cite{Vicsek2012,Chate2020}, that describe cases where alignment interactions are dominant, and Active Brownian Particles \cite{Howse2007,Romanczuk2012} or Run and Tumble particles \cite{Schnitzer1993}, that describe systems dominated by steric forces randomising their self-propulsion direction either smoothly or in a discontinuous manner.  
%For dry active matter, the Vicsek model \cite{Chate2008,Vicsek2012} has been identified as one of the key systems. 
In this work, we study a minimal model for dilute suspensions of motile bacteria that, arguably, play the same role for
wet active matter \cite{Koch2011,Saintillan2013}.

Collective motion in bacteria has been extensively studied in dilute \cite{Soni2003,Dombrowski2004,Gachelin2014} and dense \cite{Mendelson1999,Wu2006,Sokolov2007,Sokolov2009,Cisneros2011,Sokolov2012,Wensink2012a,Dunkel2013} suspensions. These studies reveal the following sequence of dynamical states. At very low densities, bacterial suspensions appear featureless and disordered \cite{Wu2006,Gachelin2014}. At higher, yet still, sufficiently low densities, collective motion sets in on the scale of the system. In this state, bacterial motion takes the form of large-scale jets and vortices with typical speeds that are larger than the swimming speeds of individual organisms \cite{Soni2003,Dombrowski2004,Gachelin2014}. At significantly higher densities, there emerges a typical lengthscale of the vortices, which is comparable to about $5-10$ times the bacterial size \cite{Sokolov2012,Dunkel2013,Ryan2013}. Although this sequence of dynamical states has never been simultaneously observed in a single systematic bulk experiment, with the exception of Sokolov \emph{et al.} \cite{Sokolov2009}, the transition scenario is supported by computer simulations of self-propelled particles interacting through various forms of long-ranged hydrodynamic fields and short-ranged steric repulsion \cite{Hernandez-Ortiz2005,Saintillan2007,Wolgemuth2008,Underhill2008,Hernandez-Ortiz2009,Lushi2013,Lushi2014,Krishnamurthy2015,Wioland2016,Saintillan2012,Stenhammar2017,Theillard2017,Schwarzendahl2018,Bardfalvy2019,Theillard2019}.

Bulk experiments with \emph{E.coli} \cite{Gachelin2014} and \emph{B.subtilis} \cite{Dombrowski2004} show that the transition to collective motion occurs around a volume fraction of bacterial bodies of about $1-2\%$. At such densities, the typical distance between organisms is about $5-8$ times their body length, collisions are rare, and the far-field hydrodynamic interactions are thought to be dominant \cite{Koch2011,Saintillan2013}. The latter are well-described by a ``pusher''-like Stokesian dipolar field \cite{Lauga2009,Drescher2011}, generated when two point forces of equal magnitude and pointing away from each other are applied to a viscous fluid. Self-propelled pusher-like dipolar particles thus form a minimal model for dilute bacterial suspensions. 

The transition to collective motion in dilute bacterial suspensions can be understood in terms of a mean-field kinetic theory \cite{Koch2011,Saintillan2013} incorporating the minimal ingredients discussed above.  Such theory identifies re-orientation of bacteria in the velocity field created by other organisms as the key ingredient leading to a global isotropic-nematic transition. The globally ordered state is, however, linearly unstable through a long-wavelength generic instability \cite{AditiSimha2002,Marchetti2013}, and there ensue never-settling dynamics as a compromise between the two instabilities. The critical density of bacteria at the onset of collective motion is determined by the strength of their dipolar interactions, their shape, and the way individual organisms change their orientation: either by occasionally re-orienting in a random way (tumbling), or by rotational diffusion \cite{Saintillan2008,Saintillan2008a,Subramanian2009,Hohenegger2010,Krishnamurthy2015}. Typically, the critical threshold density is significantly lower in the latter case, and going to zero in the absence of a decorrelation mechanism for individual bacterium orientation. The mean-field kinetic theory has also been extended to systems with steric interactions \cite{Ezhilan2013,Ryan2013,Heidenreich2016,Reinken2018} and to microswimmers suspended in non-Newtonian fluids \cite{Bozorgi2013,Bozorgi2014,Li2016}.

Below the onset of collective motion, the mean-field kinetic theory predicts that the suspension is homogeneous and isotropic, as featureless as  a suspension of non-interacting microswimmers. These assumptions are widely used when describing rheological properties of very dilute suspensions \cite{Hatwalne2004,Chen2007,Sokolov2009a,Saintillan2010,Underhill2011,Lopez2015,Alonso-Matilla2016,Nambiar2017,Guo2018,Nambiar2019,Liu2019,Saintillan2018,Martinez2020} and enhanced diffusivity of tracer particles \cite{Wu2000,Kim2004,Underhill2008,Leptos2009,Dunkel2010,Ishikawa2010,Childress2010,Childress2011,Kurtuldu2011,Mino2011,Mino2013,Jepson2013,Pushkin2013,Pushkin2013jfm,Morozov2014,Kasyap2014,Thiffeault2015,Patteson2016,Burkholder2017}. However, recent large-scale Lattice-Boltzmann simulations of dipolar swimmers \cite{Stenhammar2017,Bardfalvy2019} revealed the presence of very strong correlations below the onset of collective motion. It was shown that various observables deviate from their mean-field values at any density of microswimmers \cite{Stenhammar2017}, with the deviation diverging in the vicinity of the onset. The origin of such strong correlations can be readily attributed to the slow spatial decay of the dipolar velocity field, implying a simultaneous coupling between all microswimmers in the system. While this argument is intuitive enough, its implementation as a theoretical framework presents major technical challenges, and only simplified cases were studied until now. In an earlier work, Underhill and Graham \cite{Underhill2011} studied the effect of correlations on the fluid velocity spatial correlation function by modelling the microswimmer orientational and positional correlations based on symmetry arguments and fixing the unknown parameters by comparing them to agent-based simulations. They reported a surprising logarithmic dependence of the fluid velocity spatial correlation function on the system size. Recent work by Nambiar \emph{et al.} \cite{Nambiar2019a} extended that result by analytically considering correlations between two microswimmers, and demonstrated that the logarithmic dependence is related to the absence of a decorrelation mechanism for microswimmer orientations, and that it disappears for run-and-tumble microswimmers. A systematic account for strong correlations between all microswimmers was achieved by Stenhammar \emph{et al.} \cite{Stenhammar2017}, who developed a kinetic theory for suspensions of ``shakers" --  particles that apply forces to the fluid but do not self-propel. A similar theory was developed by Qian \emph{et al.} \cite{Qian2017}, who studied a stochastic kinetic theory for two-dimensional suspensions of swimming microorganisms. Analytical results obtained in that work were limited to the case of slow swimming -- a perturbation theory that assumes that microswimmer self-propulsion is a small effect compared to their thermal diffusion and advection by other microswimmers. Such slow microswimmers are practically indistinguishable from shakers, and these results have a similar status as the theory by Stenhammar \emph{et al.} \cite{Stenhammar2017}.

In this work we develop a kinetic theory that goes beyond the mean-field assumption for the general model of dilute microswimmer suspensions described above. Our theory explicitly includes particle self-propulsion of arbitrary strength, and is valid at any density of microswimmers below the onset of collective motion. This constitutes simultaneously a significant methodological  development compared to the work by Stenhammar \emph{et al.} \cite{Stenhammar2017}, and a major advance in our understanding of one of the key models defining ``wet" active matter. Our theory allows us to make explicit predictions for observables that can be directly set against experiments and numerical simulations. 

The paper is organised as follows. In Section \ref{section:theory} we formulate a kinetic theory for a model suspension of pusher-like dipolar microswimmers. We explicitly find the dynamics of fluctuations around the homogeneous and isotropic state that describe the system below the onset of collective motion. Since our theory differs significantly from the
previous work \cite{Stenhammar2017}, we present its derivation in detail. We appreciate, however, that some readers might only be interested in the results of our theory without feeling the need to go through the rather technical Section \ref{section:theory}. We, therefore, present our results in a stand-alone Section \ref{section:results}, which can be read without Section \ref{section:theory}. There, we calculate the temporal and spatial correlation functions, fluid velocity variance, energy spectra, and the enhanced diffusivity of tracer particles. We conclude in Section \ref{section:discussion}, while Appendices contain additional derivations for technically oriented readers.

%\begin{itemize}

%\item{
%Experiments on collective motion:
%Quite dense B.S. : \cite{Mendelson1999,Sokolov2007,Sokolov2009,Cisneros2011,Sokolov2012,Wensink2012a,Dunkel2013}
%Quite dense E.coli: \cite{Wu2006}
%Dilute: \cite{Soni2003,Dombrowski2004,Gachelin2014}
%Correlation length dense B.S. \cite{Ryan2013}.
%}

%\item{
%Simulations of collective motion
%
%like us \cite{Saintillan2007,Wolgemuth2008,Underhill2008,Lushi2013,Krishnamurthy2015,Saintillan2012,Stenhammar2017,Bardfalvy2019}
%
%confinement \cite{Hernandez-Ortiz2005,Hernandez-Ortiz2009,Lushi2014,Wioland2016,Theillard2017,Theillard2019}
%}

%\item{
%Theory collective motion: \cite{Saintillan2008,Saintillan2008a,Subramanian2009,Hohenegger2010,Krishnamurthy2015}, kinetic theory with steric repulsion \cite{Ezhilan2013,Ryan2013,Heidenreich2016,Reinken2018}, viscoelastic \cite{Bozorgi2013,Bozorgi2014,Li2016}, 
%}

%\item{Generic instability: \cite{AditiSimha2002}}

%\item{
%Correlations: \cite{Stenhammar2017,Qian2017,Nambiar2019a}
%}

%\item{
%Rheology \cite{Hatwalne2004,Chen2007,Sokolov2009a,Saintillan2010,Underhill2011,Lopez2015,Alonso-Matilla2016,Nambiar2017,Guo2018,Nambiar2019,Liu2019,Saintillan2018}.
%}

%\item{
%Enhanced diffusion 
%theory:
%\cite{Underhill2008,Dunkel2010,Ishikawa2010,Childress2010,Childress2011,Pushkin2013,Pushkin2013jfm,Morozov2014,Kasyap2014,Thiffeault2015,Burkholder2017}
%exp: \cite{Wu2000,Kim2004,Leptos2009,Kurtuldu2011,Mino2011,Mino2013,Jepson2013,Patteson2016}
%}

%\item{
%Silly correlations: 
%\cite{Liao2007,Lau2009}??
%}
%
%\end{itemize}

\section{Kinetic theory of strongly interacting suspensions}
\label{section:theory}

\subsection{Microscopic model}
\label{subsection:microscopic}
We consider a collection of $N$ microswimmers contained in a volume $V$ at a finite number density $n=N/V$. The microswimmers are suspended in a Newtonian fluid with the viscosity $\mu$. 
Each microswimmer is described by its instantaneous position ${\bm x}_i$ and orientation ${\bm p}_i$, that we collectively denote by ${\bm z}_i = \left( {\bm x}_i, {\bm p}_i\right)$, where $i=1\dots N$ enumerates the particles. Within our model, 
the dynamics of the suspension is governed by the following equations of motion
\begin{align}
\dot{x}_{i}^{\alpha} &= v_{s} p_{i}^{\alpha} + \mathcal{U}^{\alpha}\left( {\bm x}_i \right),   
\label{xdot} \\
%%%%%%%%%%%%%%%%%%%%%%%%%%%%%%%%%%%%%%%%%%%%%%%%%%%%%
\dot{p}_{i}^{\alpha} &= \mathbb{P}_{i}^{\alpha\beta} \left(\mathcal{W}^{\beta\gamma}\left( {\bm x}_i \right) + B \mathcal{E}^{\beta\gamma}\left( {\bm x}_i \right) \right) p_{i}^{\gamma},
\label{pdot}
\end{align}
where the dot denotes the time derivative, the superscript indices denote Cartesian components of vectors, and the subscript indices label the particles. Throughout this work, we utilise the Einstein summation convention for superscript indices, while no summation is assumed over repeated subscript indices.

The equations of motion \eqref{xdot} and \eqref{pdot} incorporate the following physical ingredients. First of all, each swimmer self-propels with the speed $v_s$ in the direction of its orientation. To induce self-propulsion, swimmers generate long-ranged flows in the suspending fluid \cite{Lauga2009}. The superposition of these flows at the position of the $i$-th swimmer, $\mathcal{U}^{\alpha}\left( {\bm x}_i \right)$, advects that particle in addition to its self-propulsion, see Eq.\eqref{xdot}, and re-orients it according to Jeffrey's equation \eqref{pdot}. The latter describes the dynamics of a passive particle in an external flow \cite{kimkarrila}, with 
%$\mathcal{E}^{\beta\gamma}$ and $\mathcal{W}^{\beta\gamma}$ 
\begin{align}
&\mathcal{W}^{\beta\gamma}\left( {\bm x}_i \right) = \frac{1}{2} \left( \nabla^{\gamma}  \mathcal{U}^{\beta}\left( {\bm x}_i \right) - \nabla^{\beta}  \mathcal{U}^{\gamma}\left( {\bm x}_i \right) \right), \\
&\mathcal{E}^{\beta\gamma} \left( {\bm x}_i \right)= \frac{1}{2} \left(\nabla^{\gamma}  \mathcal{U}^{\beta}\left( {\bm x}_i \right) + \nabla^{\beta}  \mathcal{U}^{\gamma}\left( {\bm x}_i \right) \right),
\label{eq:Strain}
\end{align}
being the Cartesian components of the vorticity  and rate-of-strain tensors, respectively.
In Eq.\eqref{pdot}, $\mathbb{P}_{i}^{\alpha\beta} = \delta^{\alpha\beta} - p^{\alpha}_{i}p_{i}^{\beta}$, is the projection operator, $\delta^{\alpha\beta}$ denotes the Kronecker delta, $\nabla_i^\alpha = \partial/\partial x_i^\alpha$,  and  $B = \left( a^2 - 1\right)/\left( a^2 + 1\right)$ is the measure of the swimmer's nonsphericity \cite{kimkarrila} based on its aspect ratio $a$. For strongly elongated particles, $B\rightarrow 1$, while for spheres, $B=0$. Finally, each swimmer randomly changes its orientation with a rate $\lambda$, thus mimicking the run-and-tumble motion commonly exhibited by bacteria \cite{Berg1993}. We note here that we neglect the effects of rotational and translational diffusion on the particle's dynamics, and random tumbling is thus the only source of stochasticity in our model. 

The velocity field generated by a self-propelled particle sufficiently far away from its surface is often well-described by the field produced by a point dipole with the same position and orientation \cite{Lauga2009,Drescher2011}. In a dilute suspension of microswimmers, where the particles are sufficiently separated from each other, we can approximate $\mathcal{U}^{\alpha}\left( {\bm x}_i \right)$ by a sum of dipolar contributions
\begin{align}
 \mathcal{U}^{\alpha}\left( {\bm x}_i \right) = \sum_{j\neq i}^{N} u_{d}^{\alpha}({\bm x}_{i}; {\bm z}_{j}),
\end{align}
where 
\begin{align}
{\bm u}_{d}({\bm x}_{i}; {\bm z}_{j}) = \frac{\kappa}{8\pi} \left[ 3
\frac{\left( {\bm p}_j\cdot {\bm x}'\right)^2 {\bm x}' + \epsilon^2 \left( {\bm p}_j\cdot {\bm x}'\right) {\bm p}_j }{\left( x'^2 + \epsilon^2 \right)^{5/2}} \right. \nonumber \\
\left. -\frac{{\bm x}'}{\left( x'^2 + \epsilon^2 \right)^{3/2}} \right]
\label{realspacedipole}
\end{align}
is the velocity field generated at ${\bm x}_{i}$ by a hydrodynamic dipole located at ${\bm x}_{j}$ with the orientation ${\bm p}_{j}$.  Here, $\kappa = F l/\mu$ is the \emph{dipolar strength}, where $F$ is the magnitude of the forces applied to the fluid, $l$ is the dipolar length, and $\mu$ is the viscosity of the fluid; ${\bm x}' = {\bm x}_i -{\bm x}_j$, and $x'$ denotes the length of ${\bm x}'$. The dipole consists of two regularised Stokeslets, that were introduced by Cortez \emph{et al.} \cite{Cortez2005}, with $\epsilon$ being the regularisation length of the order of swimmer size. For pushers, $\kappa > 0$.
{\alex For free-swimming \emph{E.coli}, the dipolar strength was measured \cite{Drescher2011} to be about $\kappa\sim800~\mu$m$^3$/s.}

The main goal of our work is to calculate spatial and temporal correlations of the fluid velocity in microswimmer suspensions described by the model above. Both quantities can be succinctly expressed through a combined correlation function
\begin{align}
&C(R,T) \nonumber \\
&\qquad = \lim_{t\rightarrow\infty} \frac{1}{V}\int d{\bm x} \,\overline{U^{\alpha}\left( {\bm x}, t \right) U^{\alpha}\left( {\bm x} + {\bm R}, t+T \right)},
\label{CRTgeneral}
\end{align} 
where $U^{\alpha}\left( {\bm x}, t \right)$ is the fluid velocity at the position ${\bm x}$ at time $t$, and the large-$t$ limit guarantees independence of the initial conditions. The spatial and temporal correlation functions are trivially recovered by setting $T=0$ and $R=0$, respectively. The bar in Eq.\eqref{CRTgeneral} denotes the average over the history of tumble events, and reflect the stochastic nature of our model. To calculate this and similar averages, below we formulate a kinetic theory of microswimmer suspensions based on our macroscopic model. Such theories have been extensively studied at the mean-field level \cite{Liao2007,Lau2009,Zaid2011,Underhill2011,Belan2019,Bardfalvy2019}. Here, we go beyond the mean-field approximation and explicitly take into account strong correlations between the swimmers caused by the long-range nature of their hydrodynamic fields, Eq.\eqref{realspacedipole}.

\subsection{Kinetic theory and BBGKY hierarchy}

%The microscopic model presented in the previous Section is stochastic in nature and calculating macroscopic quantities characterising the system requires taking averages over the initial conditions and the history of tumble event. To this end, we formulate here a kinetic theory of a microswimmer suspension taking into account strong correlations between the swimmers caused by the long-range nature of their hydrodynamic fields, Eq.\eqref{realspacedipole}.

The starting point of our theory is the $N$-particle probability distribution function $F_N\left({\bm z}_1, {\bm z}_2, \dots, {\bm z}_N, t \right)$ that gives the geometric probability of the system occupying a particular point in the $6N$-dimensional phase space $\left\{ {\bm z}_1, \dots, {\bm z}_N\right\} $ at time $t$. The $N$-particle probability distribution function is symmetric with respect to swapping particle labels, reflecting their indistinguishability, and is normalised
\begin{align}
\int d{\bm z}_1 \dots d{\bm z}_N F_N\left({\bm z}_1, \dots, {\bm z}_N, t \right) = 1.
\label{normalisation}
\end{align}
Its time dynamics is governed by the Master equation \cite{Balescu1975}
\begin{align}
\partial_{t} F_{N} & + \sum_{i=1}^{N} \Big[ \nabla_{i}^{\alpha}(\dot{x}_{i}^{\alpha} F_{N}) + \partial_{i}^{\alpha}(\dot{p}_{i}^{\alpha} F_{N}) \Big] \nonumber \\
& = - N \lambda F_{N} +  \frac{\lambda}{4\pi} \sum_{i=1}^{N} \int \mathrm{d} {\bm p}_{i} F_{N},
\label{Liouville}
\end{align}
where we introduced $\partial_{i}^{\alpha} = \mathbb{P}_{i}^{\alpha\beta} \partial/\partial p^\beta_i$. The l.h.s. of Eq.\eqref{Liouville} describes the probability fluxes to and from a particular point in the phase space due to the deterministic particle dynamics given by Eqs.\eqref{xdot} and \eqref{pdot}, while the r.h.s. gives the changes of the probability due to random tumbling from and into that phase space point \cite{Subramanian2009,Koch2011}. Next, we introduce the $s$-particle correlation functions defined as
\begin{align}
&F_{s}\left({\bm z}_1,\dots, {\bm z}_s, t \right) = \frac{N!}{(N-s)!N^{s}} \nonumber \\
&\qquad\qquad \times \int d{\bm z}_{s+1} \dots d{\bm z}_N F_N\left({\bm z}_1, \dots, {\bm z}_N, t \right), 
\label{Fs}
\end{align}
Below, we will only be interested in the first partial correlation functions $F_1$, $F_2$, and $F_3$, that we further express as
\begin{align}
F_2\left( {\bm z}_1, {\bm z}_2, t\right) = F_1\left( {\bm z}_1, t\right) F_1\left( {\bm z}_2, t \right) +  G\left( {\bm z}_1, {\bm z}_2, t\right),
\label{F2}
\end{align}
and
\begin{align}
F_3\left( {\bm z}_1, {\bm z}_2, {\bm z}_3, t\right) = F_1\left( {\bm z}_1, t\right) F_1\left( {\bm z}_2, t \right) F_1\left( {\bm z}_3, t \right) \nonumber \\
+ G\left( {\bm z}_1, {\bm z}_2, t\right) F_1\left( {\bm z}_3, t \right) + G\left( {\bm z}_1, {\bm z}_3, t\right) F_1\left( {\bm z}_2, t \right) \nonumber \\
+ G\left( {\bm z}_2, {\bm z}_3, t\right) F_1\left( {\bm z}_1, t \right) + H\left( {\bm z}_1, {\bm z}_2, {\bm z}_3, t\right),
\label{F3}
\end{align}
where $G$ and $H$ are the irreducible (connected) correlation functions \cite{Balescu1975}. The time evolution of $F_s$ can be deduced from the Master equation \eqref{Liouville} by integrating it over $\left\{{\bm z}_{s+1},\dots,{\bm z}_N\right\}$. Integrating by parts and using Eqs.\eqref{F2} and \eqref{F3}, we obtain the following equations for the one- and two-particle irreducible correlation functions
\begin{widetext}

\begin{align}
& \partial_t F_1({\bm z}, t) + {\mathcal L}[F_1({\bm z}, t)]({\bm z}) \nonumber \\
& \qquad\qquad = -N \nabla^\alpha \int d{\bm z}' G({\bm z},{\bm z}',t) u^\alpha_d ({\bm x}; {\bm z}')  
- N \mathbb{P}^{\alpha\beta} \frac{\partial}{\partial p^\beta} \int d{\bm z}' G({\bm z},{\bm z}',t)
p^\gamma \mathbb{X}^{\alpha\mu\nu\gamma} \nabla^\mu u^\nu_d ({\bm x}; {\bm z}'), 
\label{EqF1}
\\ 
\nonumber \\
%%%%%%%%%%%%%%%
%%%%%%%%%%%%%%%
& \partial_t G({\bm z}_1,{\bm z}_2, t) + {\mathcal L}[G({\bm z}_1,{\bm z}_2, t)]({\bm z}_1) + {\mathcal L}[G({\bm z}_1,{\bm z}_2, t)]({\bm z}_2) \nonumber \\
%%%% Term1 Particle 1 %%%%%%%
& \qquad \qquad + N \nabla_1^\alpha \left[ F_1({\bm z}_1,t) \int d{\bm z}' G({\bm z}_2,{\bm z}',t) u_d^\alpha({\bm x}_1;{\bm z}')\right]
%%%% Term1 Particle 1 %%%%%%%
+ N \nabla_2^\alpha \left[ F_1({\bm z}_2,t) \int d{\bm z}' G({\bm z}_1,{\bm z}',t) u_d^\alpha({\bm x}_2;{\bm z}')\right]
\nonumber \\ 
%%%% Particle 1 %%%%%%%
& \qquad \qquad + N \mathbb{P}_1^{\alpha\beta} \frac{\partial}{\partial p_1^\beta}\left[  F_1({\bm z}_1,t) p_1^\gamma \mathbb{X}_1^{\alpha\mu\nu\gamma} \int d{\bm z}' G({\bm z}_2,{\bm z}',t)\nabla_1^\mu u_d^\nu({\bm x}_1;{\bm z}')\right]
\nonumber \\ 
%%%% Particle 2 %%%%%%%
& \qquad \qquad + N \mathbb{P}_2^{\alpha\beta} \frac{\partial}{\partial p_2^\beta}\left[  F_1({\bm z}_2,t) p_2^\gamma \mathbb{X}_2^{\alpha\mu\nu\gamma} \int d{\bm z}' G({\bm z}_1,{\bm z}',t)\nabla_2^\mu u_d^\nu({\bm x}_2;{\bm z}')\right] \nonumber \\
& \qquad \qquad =  - \mathcal{S}_{1,2}^{F} - \mathcal{S}_{2,1}^{F} -  \mathcal{S}_{1,2}^{G} - \mathcal{S}_{2,1}^{G} - \mathcal{S}_{1,2}^{H} - \mathcal{S}_{2,1}^{H},
\label{EqG}
\end{align}
where we have introduced the operator
\begin{align}
&{\mathcal L}[\Phi]({\bm z}) \nonumber \\
& \qquad =
v_s p^\alpha \nabla^\alpha \Phi({\bm z}) 
+ N \nabla^\alpha \big[ \Phi({\bm z}) \mathcal{U}^\alpha_{\text{MF}}({\bm x})\big] 
+ N \mathbb{P}^{\alpha\beta} \frac{\partial}{\partial p^\beta}\big[ \Phi({\bm z}) p^\gamma \mathbb{X}^{\alpha\mu\nu\gamma} \nabla^\mu \mathcal{U}^\nu_{\text{MF}}({\bm x})\big] 
+ \lambda \Phi({\bm z})
- \frac{\lambda}{4\pi} \int d{\bm p}\, \Phi({\bm z}),
\label{operatorL}
\end{align}

\end{widetext}
acting on the variable $\bm z$ of an arbitrary function $\Phi=\Phi({\bm z}_1,\dots,{\bm z}_N)$, and defined the mean-field velocity field as
\begin{align}
 \mathcal{U}^\alpha_{\text{MF}}({\bm x}) = \int d{\bm z}' F_1({\bm z}',t) u^\alpha_d({\bm x}; {\bm z}').
 \label{UMF}
\end{align}
The rank-4 tensor
\begin{align}
\mathbb{X}_i^{\alpha\mu\nu\gamma} = \mathbb{P}_i^{\alpha\beta} \left[ \frac{B+1}{2}\delta^{\mu\gamma}\delta^{\nu\beta} + \frac{B-1}{2}\delta^{\mu\beta}\delta^{\nu\gamma} \right],
\end{align}
encodes the tensorial structure of Jeffrey's equation \eqref{pdot}, and the r.h.s. of Eq.\eqref{EqG} is given in terms of
\begin{align} %%%%%% Sf %%%%%%%
&\mathcal{S}_{i,j}^{F} = 
F_1({\bm z}_j,t)\Big\{ \nabla_i^\alpha \left[ F_1({\bm z}_i,t) u_d^\alpha({\bm x}_i;{\bm z}_j)\right] \nonumber \\
&+  \mathbb{P}_i^{\alpha\beta} \frac{\partial}{\partial p_i^\beta}\left[ F_1({\bm z}_i, t) p_i^\gamma \mathbb{X}_i^{\alpha\mu\nu\gamma} \nabla_i^\mu u_d^\nu({\bm x}_i; {\bm z}_j) \right] \Big\},
\end{align}
\begin{align} %%%%%% Sg %%%%%%%
& \mathcal{S}_{i,j}^{G} = 
\nabla_i^\alpha \left[ G({\bm z}_i,{\bm z}_j,t) u_d^\alpha({\bm x}_i;{\bm z}_j)\right] \nonumber \\
&+  \mathbb{P}_i^{\alpha\beta} \frac{\partial}{\partial p_i^\beta}\left[ G({\bm z}_i,{\bm z}_j,t) p_i^\gamma \mathbb{X}_i^{\alpha\mu\nu\gamma} \nabla_i^\mu u_d^\nu({\bm x}_i; {\bm z}_j) \right], 
\end{align}
and
\begin{align} %%%%%% ST %%%%%%%
& \mathcal{S}_{i,j}^{H} = 
N \int d{\bm z}' \Big\{
\nabla_i^\alpha \left[ H({\bm z}_i,{\bm z}_j,{\bm z}',t) u_d^\alpha({\bm x}_i;{\bm z}')\right] \nonumber \\
&+  \mathbb{P}_i^{\alpha\beta} \frac{\partial}{\partial p_i^\beta}\left[ H({\bm z}_i,{\bm z}_j,{\bm z}',t) p_i^\gamma \mathbb{X}_i^{\alpha\mu\nu\gamma} \nabla_i^\mu u_d^\nu({\bm x}_i; {\bm z}') \right] \Big\}.
\end{align}
Eqs.\eqref{EqF1} and \eqref{EqG} are the beginning of a BBGKY hierarchy of equations for partial distribution functions \cite{Balescu1975}. As such, they do not form a closed system as they also depend on the three-particle irreducible distribution function $H$. {\alex The BBGKY equations have been extensively studied before \cite{Fisher1961,Balescu1975}, and they form one of the main tools of analysing statistical properties of many-body systems. Here, we develop a similar technique for a collection of microswimmers with long-range hydrodynamic interactions. The assumptions we make below are based upon the previous literature on BBGKY equations in systems with long-range interactions \cite{Nicholson1983,Campa2009,Heyvaerts2010,Nardini2012a,Nardini2012b}, such as plasmas and self-gravitating matter.}

Before discussing our choice of closure for this system of equations, let us briefly review the predictions of the mean-field approximation to Eqs.\eqref{EqF1} and \eqref{EqG}, which consists of neglecting all correlation functions beyond $s=1$. The remaining equation determines the mean-field approximation to the one-particle correlation function
\begin{align}
\partial_t F_1^{\text{MF}}({\bm z}, t) + {\mathcal L}[F_1^{\text{MF}}({\bm z}, t)]({\bm z})  = 0,
\label{EqF1MF}
\end{align}
that has been extensively studied before \cite{Saintillan2008,Saintillan2008a,Subramanian2009,Hohenegger2010,Koch2011,Saintillan2013,Krishnamurthy2015}.
One of the solutions of this equation is given by a constant, which is fixed to $F_1^{\text{MF}}({\bm z}, t)=1/(4\pi V)$ by the normalisation condition Eq.\eqref{normalisation}. This solution, which is valid at any number density, corresponds to a homogeneous and isotropic suspension of microswimmers. 
For pushers $(\kappa>0)$, this state loses its stability \cite{Saintillan2008, Saintillan2008a,Subramanian2009,Hohenegger2010,Stenhammar2017} at the critical number density of microswimmers $n_{crit} = 5\lambda/(B\kappa)$, while for pullers $(\kappa<0)$, the homogeneous and isotropic state is always linearly stable within the mean-field approximation. 

The homogeneous and isotropic mean-field solution implies that $N F_1^{\text{MF}}\sim n \sim O(1)$ is finite in the thermodynamic limit. 
%While higher-order correlators with $s>1$ can potentially impose small corrections to this behaviour, Eq.\eqref{EqF1} suggests that to leading order $N F_1 \sim O(1)$, even in the presence of correlations. 
This, in turn, implies that, to leading order, $G\sim O(N^{-2})$, $H\sim O(N^{-3})$, etc. A more comprehensive discussion of this statement, together with the required rescaling of the correlation functions, system parameters, and time is given elsewhere \cite{Stenhammar2017}. 

Building upon these results, here we \emph{assume} that upon approaching the thermodynamic limit, $F_1$ is well-approximated by $F_1^{\text{MF}}$, since the r.h.s. of Eq.\eqref{EqF1} is $O(1/N)$ compared to its l.h.s. In the homogeneous and isotropic state, the mean-field velocity vanishes $\mathcal{U}^\alpha_{\text{MF}}({\bm x}) = 0$, since the integral in Eq.\eqref{UMF} is then proportional to the total flow rate through a surface surrounding the dipole. The latter is zero due to incompressibility. {\alex Fluctuations around the homogeneous and isotropic state are then governed by Eq.\eqref{EqG} with $F_1=1/(4\pi V)$. In the thermodynamic limit, $\mathcal{S}_{i,j}^{G}$ and $\mathcal{S}_{i,j}^{H}$ are small compared to $\mathcal{S}_{i,j}^{F}$, and will thus be neglected. Effectively, this \emph{weak-coupling approximation} \cite{Campa2009,Heyvaerts2010} ignores the three-point irreducible correlations $H$. The resulting equation reads}
\begin{align}
& \partial_t G({\bm z}_1,{\bm z}_2, t) + {\mathcal L}_{12} [G] + {\mathcal L}_{21} [G] \nonumber \\
& \qquad\qquad = \frac{3B}{\left(4\pi V\right)^2} \Big\{ p_1^\mu p_1^\nu \nabla_1^\mu u_d^\nu({\bm x}_1;{\bm z}_2) \nonumber \\
& \qquad\qquad\qquad\qquad\qquad+ p_2^\mu p_2^\nu \nabla_2^\mu u_d^\nu({\bm x}_2;{\bm z}_1) \Big\},
\label{EqGfinal}
\end{align}
where
\begin{align}
& {\mathcal L}_{ij} [G] = 
v_s p_i^\alpha \nabla_i^\alpha G({\bm z}_1,{\bm z}_2, t) \nonumber \\
&\qquad\qquad - \frac{3 n B}{4\pi} p_i^\mu p_i^\nu \int d{\bm z}' G({\bm z}_j,{\bm z}', t) \nabla_i^\mu u_d^\nu({\bm x}_i;{\bm z}') \nonumber \\
& \qquad\qquad + \lambda G({\bm z}_1,{\bm z}_2, t)
- \frac{\lambda}{4\pi} \int d{\bm p}_i\, G({\bm z}_1,{\bm z}_2, t).
\label{Lij}
\end{align}
{\alex This equation has a transparent physical interpretation. First, correlations between two particles are generated by their mutual re-orientation, as encoded in the r.h.s of Eq.\eqref{EqGfinal}. Next, correlations are changed by each particle's self-propulsion and tumbling, represented by the first, and third and fourth terms in Eq.\eqref{Lij}, respectively. Finally, each particle in the pair is re-oriented by the velocity field created by all other particles that are correlated with the second particle in the pair. Effectively, this term renormalises the strength of the forcing on the r.h.s. of Eq.\eqref{EqGfinal}, and is reminiscent of the renormalisation techniques developed in sedimentation \cite{Batchelor1972,Hinch1977}. Remarkably, owing to the fact that $\mathcal{U}^\alpha_{\text{MF}}({\bm x}) = 0$, Eq.\eqref{EqGfinal} does not contain the effect of mutual advection by microswimmers underscoring the purely orientational origin of their correlations.}

{\alex Eq.\eqref{EqGfinal} has previously been derived and analysed for the case of shakers ($v_s=0$) \cite{Stenhammar2017}. We will now proceed to solve it in the general case $v_s>0$.}

\subsection{Phase-space density fluctuations}

While the two-point distribution function $G$, given by Eq.\eqref{EqGfinal}, contains statistical information about fluctuations in the system, it is not straightforward to relate it to the spatial and temporal correlation function $C(R,T)$, Eq.\eqref{CRTgeneral}, that we seek to calculate. To establish this connection, we introduce a method based on the phase space density
\begin{align}
\varphi({\bm z},t) = \sum_{i=1}^N \delta({\bm z} - {\bm z}_i(t)),
\end{align}
pioneered by Klimontovich \cite{klimontovich1967book}. Here, $\delta({\bm z})$ is the three-dimensional Dirac delta function. The average of the phase space density is related to $F_1$ as can be seen from
\begin{align}
&\overline\varphi({\bm z},t) = \int d{\bm z}_1 \dots d{\bm z}_N  \sum_{i=1}^N \delta({\bm z} - {\bm z}_i) F_N({\bm z}_1, \dots, {\bm z}_N, t) \nonumber \\
& \qquad\qquad = N F_1({\bm z}, t),
\end{align}
where we used Eq.\eqref{Fs}. Fluctuations of the phase space density can formally be defined as $\delta \varphi = \varphi - \overline\varphi$, and their second moment is given by
\begin{align}
& G_K( {\bm z}', {\bm z}'', t) \equiv \overline{\delta \varphi({\bm z}',t) \delta \varphi({\bm z}'',t)} \nonumber \\
& \qquad = N^2 G ( {\bm z}', {\bm z}'', t) + N F_1({\bm z}',t) \delta({\bm z}' - {\bm z}'').
\label{GK}
\end{align}
Below, we refer to $G_K$ as the Klimontovich correlation function. Its utility is evident if one considers the spatial correlation function $C(R)$, defined in Eq.\eqref{CRTgeneral} as
\begin{align}
&C(R) = \lim_{t\rightarrow\infty} \frac{1}{V}\int d{\bm x} \overline{U^{\alpha}\left( {\bm x}, t \right) U^{\alpha}\left( {\bm x} + {\bm R}, t \right)}.
\end{align} 
The velocity of the fluid at a position $\bm x$ is given by the superposition of the velocity fields generated by all swimmers
\begin{align}
&U^{\alpha}\left( {\bm x}, t \right) = \sum_{i=1}^N u_d^\alpha({\bm x}; {\bm z}_i(t)) \nonumber \\
&\qquad\qquad = \int d{\bm z}' \varphi({\bm z}',t) u_d^\alpha({\bm x}; {\bm z}'). 
\end{align}
Separating the phase space density into its average and fluctuations, $\varphi =\overline\varphi +  \delta \varphi$, the spatial correlation function becomes
\begin{align}
&C(R) = \lim_{t\rightarrow\infty} \frac{1}{V}\int d{\bm x} \int d{\bm z}'d{\bm z}''  u_d^\alpha({\bm x}; {\bm z}') u_d^\alpha({\bm x}+{\bm R}; {\bm z}'')  \nonumber \\
& \qquad\qquad\qquad \times  \Bigg[ \left(\frac{n}{4\pi}\right)^2 + G_K( {\bm z}', {\bm z}'', t)\Bigg].
\end{align}
The integral with the constant term vanishes, demonstrating that $G_K$ fully determines the spatial correlation function. 

Time evolution of the Klimontovich correlation function can readily be derived from Eqs.\eqref{EqGfinal} and \eqref{GK}, yielding
\begin{align}
& \partial_t G_K({\bm z}_1,{\bm z}_2, t) + {\mathcal L}_{12} [G_K] + {\mathcal L}_{21} [G_K] \nonumber \\
& \qquad\qquad = 
2 \lambda \frac{n}{4\pi} \delta({\bm x}_1 - {\bm x}_2 ) \left[ \delta({\bm p}_1 - {\bm p}_2 ) - \frac{1}{4\pi} \right],
\label{EqGK}
\end{align}
where $ {\mathcal L}_{ij}$ is defined in Eq.\eqref{Lij}, and we used $F_1 = 1/(4\pi V)$ in the homogeneous and isotropic state. To solve Eq.\eqref{EqGK}, we introduce an auxiliary field $h({\bm z}_1, t)$, that satisfies the following equation
\begin{align}
 \partial_t h({\bm z}_1, t) + {\mathcal L}_{11} [h] = \chi({\bm z}_1, t),
\label{Eqh}
\end{align}
where $\chi$ is a noise term with the following properties
\begin{align}
& \langle \chi({\bm z}_1, t) \rangle = 0, 
\label{noiseaverage}\\
& \langle \chi({\bm z}_1, t) \chi({\bm z}_2, t')\rangle = 2 \lambda \frac{n}{4\pi} \delta(t-t') \delta({\bm x}_1 - {\bm x}_2 ) 
\nonumber \\
&\qquad\qquad \times \left[ \delta({\bm p}_1 - {\bm p}_2 ) - \frac{1}{4\pi} \right].
\label{noisevariance}
\end{align}
Here, the angular brackets denote the average over the realisations of the noise $\chi$, and should not be confused with the ensemble averages that we denoted by bars in the equations above. Eq.\eqref{Eqh} allows us to factorise the Klimontovich correlation function as
\begin{align}
G_K({\bm z}_1,{\bm z}_2, t) = \langle h({\bm z}_1, t) h({\bm z}_2, t)\rangle,
\end{align}
which replaces the deterministic Eq.\eqref{EqGK} by a significantly simpler stochastic Eq.\eqref{Eqh} with a fictitious noise $\chi$ with properly chosen spectral properties. Remarkably, the non-equal time correlations of the phase space density can be expressed through the same auxiliary field
\begin{align}
\overline{\delta \varphi({\bm z}',t') \delta \varphi({\bm z}'',t'')} = \langle h({\bm z}', t') h({\bm z}'', t'')\rangle,
\end{align}
as implied by a seminal work of Klimontovich and Silin \cite{Silin1962}. This, finally, leads to a direct relationship between the field $h$, which encodes the statistical properties of fluctuations in the suspension, and the combined correlation function
\begin{align}
&C(R,T) = \lim_{t\rightarrow\infty} \frac{1}{V}\int d{\bm x} \int d{\bm z}'d{\bm z}''  \nonumber \\
&  \times u_d^\alpha({\bm x}; {\bm z}') u_d^\alpha({\bm x}+{\bm R}; {\bm z}'')  \langle h({\bm z}', t) h({\bm z}'', t+T) \rangle.
\label{CRTtmp}
\end{align} 

\subsection{Dynamics of the auxiliary field $h$}
\label{subsection:h}

Here, we explicitly find the solution to Eq.\eqref{Eqh} together with Eqs.\eqref{noiseaverage} and \eqref{noisevariance}. Since Eq.\eqref{Eqh} is linear in $h$, we introduce the Fourier
\begin{align}
h({\bm z}, t) = \frac{1}{(2\pi)^3} \int d{\bm k} e^{i {\bm k}\cdot{\bm x}} \hat{h}({\bm k}, {\bm p}, t),
\label{FT}
\end{align}
and the Laplace transforms
\begin{align}
\hat{h}({\bm k}, {\bm p}, s) = \int_0^\infty dt e^{-s t} \hat{h}({\bm k}, {\bm p}, t).
\end{align}
We will also require the Fourier transform of the regularised dipolar field, Eq.\eqref{realspacedipole}, which is given by
\begin{align}
& u_d^\nu({\bm x}; {\bm z}') = \frac{-i \kappa}{(2\pi)^3}  \int d{\bm k} \, e^{i {\bm k}\cdot ({\bm x}-{\bm x}')}  \nonumber \\
& \qquad\qquad\qquad \times  \frac{A(k \epsilon)}{k} (\hat{\bm k} \cdot {\bm p}') \left( \delta^{\nu\delta} - \hat{k}^\nu \hat{k}^\delta \right) p'^{\delta},
\label{udfourier}
\end{align}
where $\hat{\bm k}={\bm k}/k$, and $k=|{\bm k}|$. The function $A$, defined as
\begin{align}
A(x)=\frac{1}{2} x^2 K_2(x),
\end{align}
with $K_2(x)$ being the modified Bessel function of the second kind, is close to unity for $x<1$, and quickly approaches zero for $x>1$. It will serve as a regularisation of the integrals over $k$, suppressing contributions from lengthscales smaller than the size of individual microswimmers. 

Performing the Fourier and Laplace transforms of Eq.\eqref{Eqh}, we obtain after re-arranging
\begin{align}
&\hat{h}({\bm k}, {\bm p}, s) = \frac{1}{\sigma({\bm k},{\bm p}, s)}
\Bigg[\hat{h}_0({\bm k}, {\bm p}) + \hat{\chi}({\bm k}, {\bm p},s) \nonumber \\
& \qquad + \frac{\lambda}{4\pi} I^{(0)}({\bm k},s)
+ \frac{15 \lambda}{4\pi} \Delta A(k\epsilon) \bigg\{ (\hat{\bm k}\cdot {\bm p}) I^{(1)}({\bm k},{\bm p}, s) \nonumber \\
& \qquad\qquad\qquad\qquad\qquad - (\hat{\bm k}\cdot {\bm p})^2 I^{(2)}({\bm k}, s)
\bigg\} 
\Bigg].
\label{hfullsolution}
\end{align} 
Here, $\hat{\chi}({\bm k}, {\bm p},s)$ is the Fourier-Laplace transform of the noise, $\sigma({\bm k},{\bm p}, s) = s + \lambda + i v_s ({\bm k}\cdot{\bm p})$, and we defined
\begin{align}
I^{(0)}({\bm k},s) &= \int d{\bm p } \,\hat{h}({\bm k}, {\bm p}, s), \label{I0} \\
I^{(1)}({\bm k},{\bm p}, s) &= \int d{\bm p}'  (\hat{\bm k}\cdot {\bm p}')({\bm p}\cdot {\bm p}') \hat{h}({\bm k}, {\bm p}', s), \label{I1} \\
I^{(2)}({\bm k},s) &= \int d{\bm p }\, (\hat{\bm k}\cdot {\bm p})^2 \hat{h}({\bm k}, {\bm p}, s). \label{I2}
\end{align}
In Eq.\eqref{hfullsolution}, $\hat{h}({\bm k}, {\bm p}, t=0) = \hat{h}_0({\bm k}, {\bm p})$ denotes some arbitrary initial condition; below we demonstrate that the long-time statistical properties of the suspension are insensitive to $\hat{h}_0({\bm k}, {\bm p})$. In Eq.\eqref{hfullsolution}, we have also introduced an important dimensionless parameter $\Delta = n/n_{crit}$, where $n_{crit} = 5\lambda/(B\kappa)$ is the mean-field onset of collective motion in pusher suspensions, $\kappa>0$. For pushers, $\Delta$ measures the dimensionless distance from the onset, with $\Delta=1$ corresponding to the instability. %For pullers, $\Delta$ is negative, with the point $\Delta=-1$ bearing no special significance.

Eq.\eqref{hfullsolution} is a linear integral equation for $\hat{h}({\bm k}, {\bm p}, s)$ and its solution is straightforward. 
Substituting  Eq.\eqref{hfullsolution} into Eqs.\eqref{I0}-\eqref{I2}, gives
\begin{align}
&I^{(0)}({\bm k},s) \nonumber \\
&\qquad\qquad= \frac{1}{1-\frac{\lambda}{4\pi}f_0} \int d{\bm p} \frac{\hat{h}_0({\bm k}, {\bm p}) + \hat{\chi}({\bm k}, {\bm p},s)}{\sigma({\bm k},{\bm p}, s)},
\label{I0sol}
\end{align}
\begin{align}
&I^{(2)}({\bm k},s) = \frac{\lambda}{4\pi} f_1 I^{(0)}({\bm k},s) \nonumber \\
& \qquad\qquad + \int d{\bm p} (\hat{\bm k}\cdot {\bm p})^2\frac{\hat{h}_0({\bm k}, {\bm p}) + \hat{\chi}({\bm k}, {\bm p},s)}{\sigma({\bm k},{\bm p}, s)},
\label{I1sol}
\end{align}
\begin{align}
&I^{(1)}({\bm k},{\bm p}, s) = \frac{1}{1+\frac{15 \lambda}{8\pi} \Delta A(k\epsilon) (f_2 - f_1)} \Bigg[ \nonumber \\
& \quad \int d{\bm p}' (\hat{\bm k}\cdot {\bm p}')({\bm p}\cdot {\bm p}') \frac{\hat{h}_0({\bm k}, {\bm p}') + \hat{\chi}({\bm k}, {\bm p}',s)}{\sigma({\bm k},{\bm p}', s)} \nonumber \\
& \quad + (\hat{\bm k}\cdot {\bm p}) \bigg\{ \frac{\lambda}{4\pi} f_1 I^{(0)}({\bm k},s) \nonumber \\
& \qquad\qquad\qquad + \frac{15 \lambda}{8\pi} \Delta A(k\epsilon) (f_2 - f_1) I^{(2)}({\bm k},s)
\bigg\} \Bigg],
\label{I2sol}
\end{align}
where
\begin{align}
f_n = 2\pi \int_{-1}^{1} dx \frac{x^{2n}}{s+\lambda + i v_s k x}.
\end{align}

Having found the explicit expression for $\hat{h}({\bm k}, {\bm p},s)$, we proceed to calculate the combined correlation function, Eq.\eqref{CRTtmp}. Below, we show that only a small number of terms from Eqs.\eqref{hfullsolution} and \eqref{I0sol}-\eqref{I2sol} contribute to $C(R,T)$.

\subsection{$C(R,T)$ in terms of $\hat{h}({\bm k}, {\bm p},s)$}
\label{subsection:hhat}

In what follows, it will be convenient to re-write $C(R,T)$ in terms of the Fourier and Laplace transforms of all quantities. 
Substituting Eq.\eqref{FT} into Eq.\eqref{CRTtmp}, and using the Fourier representation of the regularised dipolar field, Eq.\eqref{udfourier}, we obtain
\begin{align}
&C(R,T) = \lim_{t\rightarrow\infty} {\mathcal L}^{-1}_{s_1,t} {\mathcal L}^{-1}_{s_2,t+T} \frac{\kappa^2}{(2\pi)^3 V} \nonumber \\
&\qquad \times \int d{\bm k} e^{-i {\bm k}\cdot{\bm R}} \frac{A^2(k\epsilon)}{k^2} 
\int d{\bm p}_1 d{\bm p}_2 
(\hat{\bm k} \cdot {\bm p}_1) (\hat{\bm k} \cdot {\bm p}_2) \nonumber \\
& \qquad\qquad\qquad \times \big( \delta^{\alpha\beta} - \hat{k}^\alpha \hat{k}^\beta \big) p_1^{\beta} 
\big( \delta^{\alpha\gamma} - \hat{k}^\alpha \hat{k}^\gamma \big) p_2^{\gamma} \nonumber \\
&  \qquad\qquad\qquad \times \langle \hat{h}({\bm k}, {\bm p}_1,s_1)\hat{h}(-{\bm k}, {\bm p}_2,s_2) \rangle_{\hat{\chi}},
\label{CRTlaplace}
\end{align} 
where ${\mathcal L}^{-1}_{s,t}$ formally denotes the inverse Laplace transform from $s$ to $t$, given by the Bromwich integral \cite{Doetsch1974}. The angular brackets $\langle\dots\rangle_{\hat{\chi}}$ denote the average with the Fourier-Laplace components of the noise $\chi$, with the following spectral properties
\begin{align}
& \langle \hat{\chi}({\bm k}, {\bm p}, s) \rangle_{\hat{\chi}} = 0, 
\label{noiseaverageFL}\\
& \langle \hat{\chi}({\bm k}, {\bm p}_1, s_1) \hat{\chi}(-{\bm k}, {\bm p}_2, s_2)\rangle_{\hat{\chi}} = 2 \lambda V \frac{n}{4\pi} 
\nonumber \\
&\qquad\qquad\qquad\qquad \times \frac{1}{s_1+s_2} \left[ \delta({\bm p}_1 - {\bm p}_2 ) - \frac{1}{4\pi} \right],
\label{noisevarianceFL}
\end{align}
obtained by applying the Fourier-Laplace transform to Eqs.\eqref{noiseaverage} and \eqref{noisevariance}.
While the average in Eq.\eqref{CRTlaplace} can readily be formed using the solution for $\hat h$ found in Section \ref{subsection:h}, the result is very cumbersome. Before proceeding, we make two observations that greatly reduce the number of terms contributing to Eq.\eqref{CRTlaplace}.

First, we observe that
\begin{align}
\int d{\bm p} \, \big( \delta^{\alpha\beta} - \hat{k}^\alpha \hat{k}^\beta \big) p^{\beta} f(\hat{\bm k} \cdot {\bm p}) = 0,
\end{align}
where $f$ is an arbitrary function of $\hat{\bm k} \cdot {\bm p}$. This statement is readily demonstrated by representing $\bm p$ in spherical coordinates with $\hat{\bm k}$ selected along the $z$-axis, and performing the angular integrals component-wise. This result has profound implications for the average $\langle \hat{h}({\bm k}, {\bm p}_1,s_1)\hat{h}(-{\bm k}, {\bm p}_2,s_2) \rangle_{\hat{\chi}}$ in Eq.\eqref{CRTlaplace}. Every term in $\hat{h}({\bm k}, {\bm p}_1,s_1)$, Eq.\eqref{hfullsolution}, that only depends on ${\bm p}_1$ through its dependence on  $(\hat{\bm k} \cdot {\bm p}_1)$ does not contribute to $C(R,T)$, as its integral over ${\bm p}_1$ with the corresponding dipolar field in Eq.\eqref{CRTlaplace} vanishes. The same applies to $\hat{h}(-{\bm k}, {\bm p}_2,s_2)$. 

The second observation is related to the initial condition. All terms that involve $\hat{h}_0({\bm k}, {\bm p})$ only depend on the Laplace frequency $s$ through $1/\sigma({\bm k},{\bm p}, s)$, and their inverse Laplace transform can be readily performed before any other integration. Since the inverse Laplace transform of $1/(s+a)$ is $e^{-a t}$, where $a$ is a complex number, the dominant long-time behaviour of such terms is given by $e^{-\lambda t}$, where we ignored the subdominant oscillatory dependencies. In Eq.\eqref{CRTlaplace}, we are interested in the $t\rightarrow$ limit, and these terms also do not contribute to $C(R,T)$.

With these observations in mind, Eq.\eqref{hfullsolution} can be significantly simplified to read
\begin{align}
&\hat{h}({\bm k}, {\bm p}, s) \cong \frac{\hat{\chi}({\bm k}, {\bm p},s)}{\sigma({\bm k},{\bm p}, s)} \nonumber \\
&\qquad\qquad + \frac{ (\hat{\bm k}\cdot {\bm p}) }{\sigma({\bm k},{\bm p}, s)}  
\frac{\frac{15 \lambda}{4\pi} \Delta A(k\epsilon) }{1+\frac{15 \lambda}{8\pi} \Delta A(k\epsilon) (f_2 - f_1)} \nonumber \\
& \qquad\qquad\qquad \times \int d{\bm p}' (\hat{\bm k}\cdot {\bm p}')({\bm p}\cdot {\bm p}') \frac{\hat{\chi}({\bm k}, {\bm p}',s)}{\sigma({\bm k},{\bm p}', s)},
\label{hneeded}
\end{align} 
where $\cong$ signifies that we only kept the terms that contribute to $C(R,T)$. Now, the average $\langle \hat{h}({\bm k}, {\bm p}_1,s_1)\hat{h}(-{\bm k}, {\bm p}_2,s_2) \rangle_{\hat{\chi}}$ assumes a tractable form that can be used in Eq.\eqref{CRTlaplace}. Separating the terms independent of $\Delta$, we obtain $C(R,T) = C_0(R,T) + C_1(R,T)$. Here, 
\begin{align}
&C_0(R,T) = \frac{\lambda n \kappa^2}{16\pi^4 } \lim_{t\rightarrow\infty} {\mathcal L}^{-1}_{s_1,t} {\mathcal L}^{-1}_{s_2,t+T}  \int d{\bm k} e^{-i {\bm k}\cdot{\bm R}} \frac{A^2(k\epsilon)}{k^2} \nonumber \\
&\quad \times \int d{\bm p} 
(\hat{\bm k} \cdot {\bm p})^2 \left[1- (\hat{\bm k} \cdot {\bm p})^2\right] \frac{1}{s_1+s_2}\nonumber \\
&  \qquad \times  
\frac{1}{\lambda + s_1 + i v_s k (\hat{\bm k}\cdot {\bm p})} 
\frac{1}{\lambda + s_2 - i v_s k (\hat{\bm k}\cdot {\bm p})},
\label{C0Laplace}
\end{align} 
represents correlations in the fluid created by non-interacting swimmers. The double inverse Laplace transform in the equation above can be performed using the method outlined in Appendix \ref{appendix:Cesare}. It yields
\begin{align}
 & \lim_{t\rightarrow\infty} {\mathcal L}^{-1}_{s_1,t} {\mathcal L}^{-1}_{s_2,t+T} \frac{1}{s_1+s_2} \frac{1}{\lambda + s_1 + i v_s k (\hat{\bm k}\cdot {\bm p})} \nonumber \\
& \qquad\times \frac{1}{\lambda + s_2 - i v_s k (\hat{\bm k}\cdot {\bm p})} = \frac{e^{-\lambda T + i v_s k T (\hat{\bm k} \cdot {\bm p})}}{2\lambda}.
\label{dilt_example}
\end{align}
Performing the angular integration, we finally obtain
\begin{align}
& C_0(R,T) = \frac{n \kappa^2 e^{-\lambda T}}{\pi^2}  \int_{0}^\infty dk \frac{\sin{k R}}{k R} A^2(k\epsilon) \nonumber \\
&\quad \times
\frac{ y(12-y^2) \cos{y} - (12-5y^2)\sin{y}
}{y^5}
\Bigg\vert_{y=v_s k T}.
\label{C0final}
\end{align}

All other terms in Eq.\eqref{CRTlaplace} correspond to additional correlations generated by the hydrodynamic interactions among the swimmers, and, as such, they are dependent on the dimensionless microswimmer density $\Delta$. Performing the angular integration over ${\bm p}_1$ and ${\bm p}_2$, gives
\begin{align}
&C_1(R,T) \nonumber \\
& = \frac{2 \lambda n \kappa^2 }{15 \pi^2} \lim_{t\rightarrow\infty} {\mathcal L}^{-1}_{s_1,t} {\mathcal L}^{-1}_{s_2,t+T} \int_{0}^\infty dk \frac{\sin{k R}}{k R} A^2(k\epsilon) \nonumber \\
&\times \frac{1}{\lambda + s_1}\frac{1}{\lambda + s_2}\frac{1}{s_1 + s_2} \frac{z_1 \psi(z_1) + z_2 \psi(z_2)}{z_1+z_2}
 \nonumber \\
&\qquad \times\Bigg[ \frac{z_1 \psi(z_1)}{\omega - z_1 \psi(z_1)} + \frac{z_2 \psi(z_2)}{\omega - z_2 \psi(z_2)} \nonumber \\
&\qquad\qquad 
+ \frac{z_1 z_2 \psi(z_1)\psi(z_2)}{\left(\omega - z_1 \psi(z_1)\right)\left(\omega - z_2 \psi(z_2)\right)} \Bigg].
\label{C1laplace}
\end{align} 
Here, we introduced $\omega = v_s k/(\lambda \Delta A(k\epsilon))$, and the function $\psi(z)$, defined as
\begin{align}
\psi(z) = \frac{5}{2}\frac{3z+2z^3-3(1+z^2)\arctan{z}}{z^5},
\label{psi}
\end{align}
which is related to $f_2-f_1$ used in the previous Section. The variable $z_i=v_s k/(\lambda + s_i)$ allows us to write Eq.\eqref{C1laplace} in a compact form but hides its complex dependence on the Laplace frequencies $s_1$ and $s_2$. 
Its inverse Laplace transform is discussed below.

\subsection{Approximate double inverse Laplace transform}
\label{subsection:approximateDILT}

The integrand of Eq.\eqref{C1laplace} is not a rational function of $s_1$ and $s_2$, and we were unable to calculate its double inverse Laplace transform exactly. Instead, here we develop a rational approximation to $\psi(z)$ that will allow us to find $C_1(R,T)$ analytically.

First, we observe that if the poles of an analytic function are known, its large-$t$ behaviour is determined by the pole with the smallest negative real part \cite{Doetsch1974}. Therefore, the presence of the pole at $-\lambda$ in Eq.\eqref{C1laplace} makes all poles with real parts smaller than $-\lambda$ irrelevant in the large-$t$ limit. This reflects the fact that individual tumbling events are always a source of de-correlation between microswimmers. 

{\alex Next, we introduce the dimensionless persistence length $L=v_s/(\lambda \epsilon)$, which compares the typical runlength of a swimmer to the dipolar regularisation size, see Eq.\eqref{realspacedipole}. Although our theory is correct for any value of $L$, in this work we consider $L=0-25$, ranging from  non-swimming (shaker) particles to wild-type \emph{E.coli} bacteria (see Section \ref{section:results} for discussion).}
%Using the definition of $z$, the argument of $\psi$ can be written as $\psi(v_s k/\lambda \times 1/(1 + \tilde{s}))$, where $\tilde{s}=s/\lambda$. 
Contributions to the integrand in Eq.\eqref{C1laplace} with $k \epsilon>1$ are strongly suppressed by the regularising factor $A(k\epsilon)$, and therefore, when approximating $\psi(z)$, the relevant domain is $-\lambda<Re(s)<0$, with  $v_s k/\lambda$ not exceeding $L$.

In Appendix \ref{appendix:psia} we show that a surprisingly good approximation to $\psi(z)$ on this domain is given by
\begin{align}
\psi_a(z) = \frac{7}{7+3 z^2}.
\label{psia}
\end{align}
The simple structure of this expression allows us to deduce the pole structure of the integrand in Eq.\eqref{C1laplace}. Indeed, with $\psi(z)$ replaced by $\psi_a(z)$, and factorising
\begin{align}
\frac{1}{\omega - z \psi(z)} = \frac{7+3z^2}{3\omega\left(z-z_{+}\right)\left(z-z_{-}\right)},
\end{align}
where 
\begin{align}
z_{\pm} =\frac{7}{6\omega} \left[ 1\pm\sqrt{1-\frac{12}{7}\omega^2} \right],
\label{z012}
\end{align}
the denominators in Eq.\eqref{C1laplace} can now be written as products of linear polynomials in $s_1$ and $s_2$. It is now straightforward to perform the inverse Laplace transform of this expression using the method outlined in Appendix \ref{appendix:Cesare}. Taking the limit of $t\rightarrow\infty$, finally gives
\begin{align}
&C_1(\rho,\tau) \nonumber \\
& = e^{-\tau} \frac{n \kappa^2 }{15 \pi^2 \epsilon} \int_{0}^\infty d\xi \frac{\sin{\xi \rho}}{\xi \rho} A^2(\xi) 
\Bigg[ 
-\cos{\left(\sqrt{\frac{3}{7}}L\xi\tau\right)} \nonumber \\
&+ \frac{e^{\frac{1}{2}A(\xi) \Delta \tau}}{1-A(\xi)\Delta + \frac{3}{7}L^2\xi^2} \Bigg\{ \nonumber \\
&  \frac{2-A(\xi)\Delta + \frac{6}{7}L^2\xi^2}{2-A(\xi)\Delta}\cosh{\left(\frac{1}{2}A(\xi) \Delta \tau  \sqrt{1-\frac{12 L^2\xi^2}{7 A^2(\xi)\Delta^2}}\right)} \nonumber \\
&\qquad +\frac{\sinh{\left(\frac{1}{2}A(\xi) \Delta \tau  \sqrt{1-\frac{12 L^2\xi^2}{7 A^2\left(\xi\right)\Delta^2}}\right)}}{\sqrt{1-\frac{12 L^2\xi^2}{7 A^2\left(\xi\right)\Delta^2}}}
\Bigg\}
\Bigg],
\label{C1final}
\end{align} 
where we changed the integration variable to $\xi = k\epsilon$, and introduced the dimensionless parameters $\rho = R/\epsilon$ and $\tau = \lambda T$. In Appendix \ref{appendix:dilt}, we verify that Eq.\eqref{C1final} provides a good approximation to the long-time behaviour of Eq.\eqref{C1laplace}. 

%Eqs.\eqref{C0final} and \eqref{C1final} constitute the main result of our work. In the next Section, we explicitly work out the predictions for the spatial and temporal correlation functions and other observables, based on this result.

\section{Results}
\label{section:results}

For the benefit of the readers who have skipped Section \ref{section:theory}, we repeat our main result, which comprises an explicit expression for the combined correlation function $C(R,T)$, defined in Eq.\eqref{CRTgeneral}. It describes the steady-state correlations between the fluid velocity at two points in space separated by a distance $R$, and two instances in time separated by a time-interval $T$. Our theory includes full hydrodynamic interactions between microswimmers and is valid at any density up to the onset of collective motion. The result consists of the non-interacting part,
\begin{align}
& C_0(\rho,\tau) = \frac{n \kappa^2 e^{-\tau}}{\pi^2\epsilon}  \int_{0}^\infty d\xi \frac{\sin{\xi \rho}}{\xi \rho} A^2(\xi) \nonumber \\
&\quad \times
\frac{ y(12-y^2) \cos{y} - (12-5y^2)\sin{y}
}{y^5}
\Bigg\vert_{y=L\xi\tau}, 
\label{ResultsC0}
\end{align}
and the interacting correlation function $C_1(\rho,\tau)$, given in Eq.\eqref{C1final}. Here, $\rho = R/\epsilon$, where $\epsilon$ is a lengthscale comparable to the microswimmer size, and $\tau = \lambda T$, where $\lambda$ is the tumbling rate. {\alex The relative distance to the instability threshold is measured by $\Delta = n/n_{crit}$, which is the dimensionless number density of the particles, where $n_{crit} = 5\lambda/(B\kappa)$ is the microswimmer number density at the onset of collective motion for pusher-like microswimmers \cite{Subramanian2009,Hohenegger2010,Stenhammar2017}}; the parameter $B$ is defined after Eq.\eqref{eq:Strain}. Our theory is valid for $\Delta<1$. 

{\alex A central role in our theory is played by the dimensionless persistence length $L=v_s/(\lambda \epsilon)$ that compares the typical distance covered by a swimming microorganism between two tumble events to the dipolar regularisation lengthscale $\epsilon$.  As we will see below, the observables we consider here depend strongly on $L$ and it is therefore important to estimate its realistic values. For wild-type \emph{E.coli} bacteria, the swimming speed is strain-dependent, and we use $v_s\sim 20-25~\mu$m/s as a representative value \cite{Darnton2007,Drescher2011}, while for the tumbling rate we use $\lambda\sim 1$~s$^{-1}$ \cite{Berg1993}. The parameter $\epsilon$, which is introduced in Eq.\eqref{realspacedipole}, regularises our theory at the length-scale below which the dipolar velocity field does not approximate sufficiently well the full velocity field created by a single bacterium. 
Na{\"i}vely, one can identify $\epsilon$ with half the body length of 
\emph{E.coli}, leading to $\epsilon\sim 1~\mu$m. A more hydrodynamically sound approach is to interpret $\epsilon$ as the length of the effective hydrodynamic dipole generated by a bacterium. Drescher \emph{et al.} \cite{Drescher2011} have measured the velocity field of swimming \emph{E.coli} bacteria far away from boundaries and concluded that it is well-represented by a pair of equal and opposite forces applied to the fluid at a distance of $1.9~\mu$m apart. Identifying the cut-off distance with half of the dipolar length again gives $\epsilon\sim1~\mu$m. In this work, we consider $L=0-25$; we 
hypothesise that this range is relevant for the wild-type \emph{E.coli}. Furthermore, using the same hydrodynamic interpretation of $\epsilon$ as above, the $L=5$ case approximately corresponds to the simulations of Stenhammar \emph{et al.} \cite{Stenhammar2017}, while the $L=0$ case describes non-swimming bacteria (shakers). Ultimately, the values of $L$ suitable for a particular microorganism will have to be determined experimentally, as we discuss in Section \ref{section:discussion}. Finally, we observe that the typical values of $L$ are higher yet for non-tumbling bacteria, where the role of the main orientation decorrelation mechanism is played by the (effective) rotational diffusion.} 

The full expression, $C(\rho,\tau)=C_0(\rho,\tau)+C_1(\rho,\tau)$, given as a definite integral, constitutes the main technical result of our study. We now explicitly work out its predictions for the spatial and temporal correlation functions, and other experimentally accessible observables. When discussing their physical meaning, we are going to vary the dimensionless persistence length $L$, while keeping all the other parameters of the microswimmers fixed. We note that in reality the dipolar strength and shape of a microorganism uniquely determine its swimming speed, and hence $L$. We, however, see varying $L$ as a tool to disentangle the effects of self-propulsion (ability to change one's position in space) from the strength of the hydrodynamic disturbances it causes. In particular, we will consider two limiting cases: shakers ($L=0$) and  fast swimmers ($L\rightarrow\infty$). The former case corresponds to microswimmers that exert dipolar forces on the fluid but do not self-propel, and only change their positions due to being advected by the velocity fields created by other microswimmers \cite{Stenhammar2017}. The latter case, while obviously non-physical, is a useful tool to assess the effect of fast swimming on various quantities of interest. When studying the behaviour of the observables listed above in the vicinity of the transition to collective motion, we will fix the value of $L$, so that fast swimming should be understood as large yet finite $L$, and consider the limit $\Delta\to1$. We will not consider the opposite order of the limits. Finally, we note that the terms representing hydrodynamic interactions in Eq.\eqref{EqGK} are proportional to the swimmer's nonsphericity $B$ that enters Jeffrey equation, Eq.\eqref{pdot}. The limit of non-interacting microswimmers therefore corresponds to setting $B$ to zero, which, in turn, can be achieved by setting $\Delta=0$, while keeping $n$ finite.

\subsection{Velocity variance}

\begin{figure}
\centering
\includegraphics[width=0.8\linewidth]{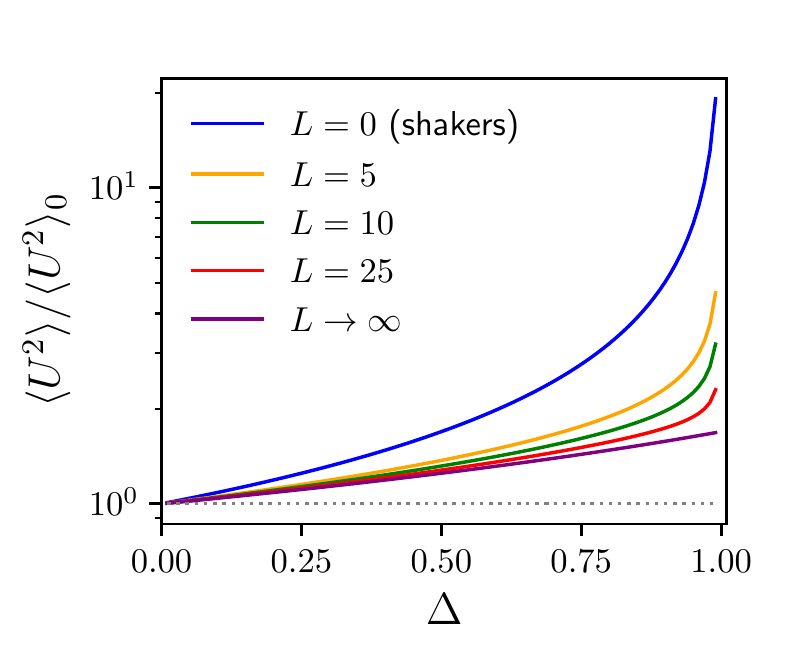}
\caption{The fluid velocity variance $\langle U^2 \rangle$ normalised by its non-interacting value $\langle U^2 \rangle_0$ for various values of $L$. The dotted line represents the non-interacting case $\langle U^2 \rangle=\langle U^2 \rangle_0$. Note that the $L\rightarrow\infty$ line turns sharply upwards and diverges in the vicinity of $\Delta=1$ in a way that cannot be resolved on the scale of this graph.}
\label{Fig_Variance}
\end{figure}

Our first quantity of interest is the fluid velocity variance, $\langle U^2 \rangle \equiv C(\rho=0,\tau=0)$. In the absence of thermal noise, re-arrangements of the microswimmer positions and orientations is the sole source of fluid velocity fluctuations. For this reason, it was used in previous studies as an order parameter to identify the onset of collective motion \cite{Stenhammar2017, Bardfalvy2019}. Summing up Eqs.\eqref{ResultsC0} and \eqref{C1final}, and setting 
$\rho=0$ and $\tau=0$,  we obtain
\begin{align}
& \langle U^2 \rangle  = \frac{\kappa^2 n}{15 \pi^2 \epsilon} \int_{0}^\infty d\xi A^2(\xi) \nonumber \\
& \qquad\qquad \times \frac{2-A(\xi)\Delta+\frac{6}{7}L^2 \xi^2}{\left(2-A(\xi)\Delta\right)\left(1-A(\xi)\Delta+\frac{3}{7}L^2 \xi^2\right)}.
\label{resultvarience}
\end{align}
We evaluate this integral numerically and plot the fluid velocity variance normalised by its value in the non-interacting case,   $\langle U^2 \rangle(\Delta=0)\equiv \langle U^2 \rangle_0$, given by \cite{Bardfalvy2019}
\begin{align}
\langle U^2 \rangle_0 = \frac{\kappa^2 n}{15 \pi^2 \epsilon} \int_{0}^\infty d\xi A^2(\xi) = \frac{21\kappa^2 n}{2048\epsilon}.
\label{eq:U2noninteracting}
\end{align}
Note that $\langle U^2 \rangle_0$ corresponds to a superposition of uncorrelated fluctuations in the fluid velocity, which, by virtue of the central limit theorem, is proportional to $n$. Any deviations of $\langle U^2 \rangle$ from that value signify the presence of correlations. 

As can be seen in Fig.\ref{Fig_Variance}, the fluid velocity fluctuations exhibit significant correlations at any density of the microswimmers, as was recognised previously \cite{Stenhammar2017}. Starting from its non-interacting value at $\Delta=0$, the variance increases with $\Delta$, until it diverges at the onset of collective motion. The strongest correlations are exhibited by suspensions of shakers, while swimming acts to reduce correlations. For large but finite values of $L$, the variance increases mildly from its non-interacting value, until it rises sharply in a small vicinity of $\Delta=1$, with the size of this region shrinking with $L$. Interestingly, the rise of $\langle U^2 \rangle_0$ for $\Delta<1$ remains finite even in the $L\rightarrow\infty$ limit. In other words, while swimming clearly reduces correlations, it does not remove them entirely, and the suspension is never described by the mean-field theory.

To determine the scaling of the fluid velocity variance as $\Delta\rightarrow 1$, we observe that in that limit the integrand in Eq.\eqref{resultvarience} is dominated by small values of $\xi$, where $A(\xi) \approx 1 - \xi^2/4$. Using this approximation in Eq.\eqref{resultvarience} and replacing the upper integration limit by unity, we obtain
\begin{align}
& \langle U^2 \rangle  \sim \frac{\kappa^2 n}{15 \pi \epsilon} \frac{1}{\sqrt{1+\frac{12}{7}L^2}\sqrt{1-\Delta}}, \quad \Delta \rightarrow 1.
\label{varianceasymptotic}
\end{align}
Therefore, our theory predicts that the fluid velocity variance diverges as $(1-\Delta)^{-1/2}$ in the vicinity of the transition to collective motion, for any finite value of $L$.

\subsection{Spatial correlations}

\begin{figure*}
\centering
\includegraphics[width=1.0\linewidth]{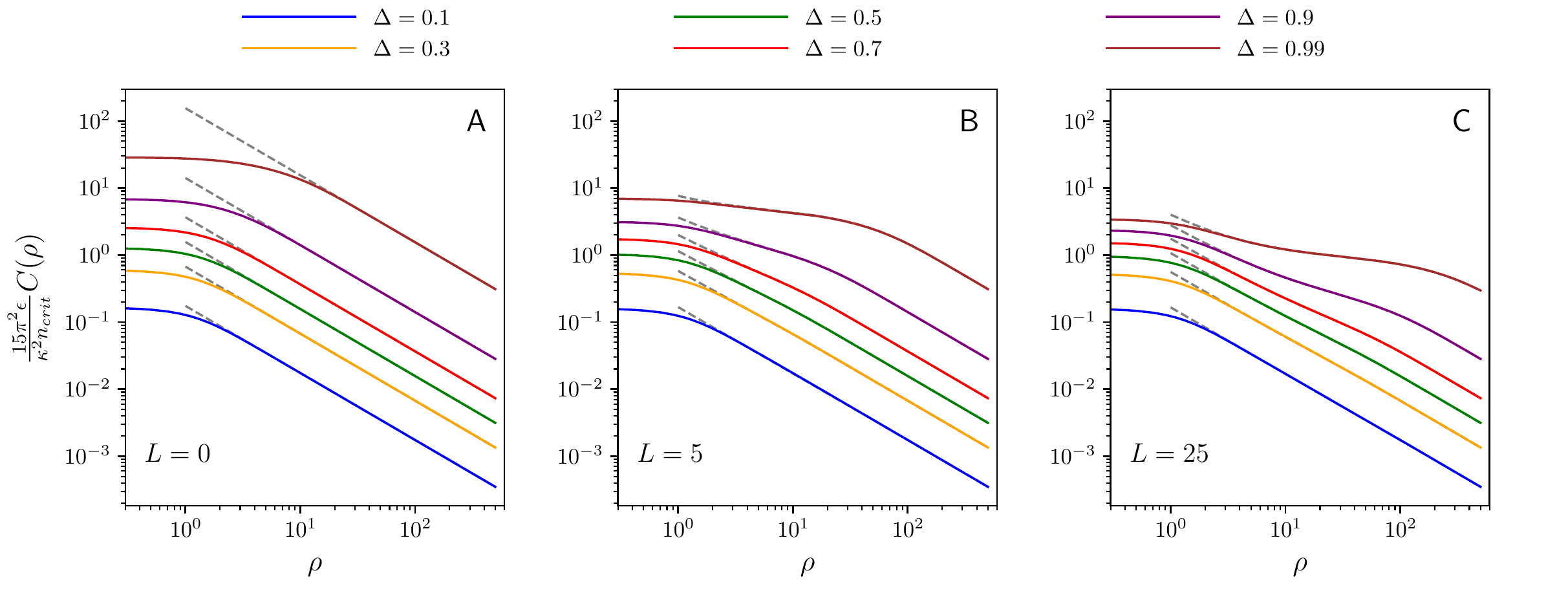}
\caption{The spatial correlation function $C(\rho)$ as a function of the distance $\rho$ for various values of $\Delta$. A: $L=0$, B: $L=5$, and C: $L=25$. The solid lines are calculated by numerically evaluating Eq.\eqref{spatialcorrelations}, while the dashed lines are the analytic approximation, Eq.\eqref{spatialcorrelationsanalytic}.  The legend applies to all panels.}
\label{Fig_spatial}
\end{figure*}

Our next quantity of interest is the equal-time spatial correlation function, $C(\rho, T=0)$, given by
\begin{align}
&C(\rho) = \frac{\kappa^2 n}{15 \pi^2 \epsilon} \int_{0}^\infty d\xi \frac{\sin{\xi \rho}}{\xi \rho} A^2(\xi) \nonumber \\
& \qquad\qquad \times \frac{2-A(\xi)\Delta+\frac{6}{7}L^2 \xi^2}{\left(2-A(\xi)\Delta\right)\left(1-A(\xi)\Delta+\frac{3}{7}L^2 \xi^2\right)}.
\label{spatialcorrelations}
\end{align}
While this integral cannot be evaluated analytically, a good approximation can be obtained by setting $A(\xi)=1$ in the integrand, yielding
\begin{align}
&C(\rho) \approx \frac{\kappa^2 n}{30 \pi \epsilon \left(1-\Delta\right)\rho} \nonumber \\
&\qquad\qquad \times \left[ 1 - \frac{\Delta}{2-\Delta} \exp{\left\{-\sqrt{\frac{7}{3}}\sqrt{1-\Delta}\frac{\rho}{L} \right\}}\right].
\label{spatialcorrelationsanalytic}
\end{align}
For $\Delta=0$, this equation reproduces the result obtained previously for non-interacting swimmers \cite{Zaid2011,Underhill2011,Belan2019,Bardfalvy2019}.

In Fig.\ref{Fig_spatial} we evaluate Eq.\eqref{spatialcorrelations} numerically and compare it against the analytic approximation, Eq.\eqref{spatialcorrelationsanalytic}; $\kappa^2 n_{crit}/(15 \pi^2 \epsilon)$
is chosen as the normalisation factor. For all values of $L$ and $\Delta$, the approximation works well for all but small spatial separations $\rho$, where the spatial correlation function is, essentially, equal to the fluid velocity variance. As with the fluid velocity variance, the strongest correlations are exhibited by suspensions of shakers, $L=0$. In this case, the spatial correlation function changes very slowly at short distances, and decays as $\rho^{-1}$ at large distances. Close to the onset of collective motion, the typical scale $\rho_0$ at which the crossover occurs can be estimated from Eqs.\eqref{varianceasymptotic} and \eqref{spatialcorrelationsanalytic}, by requiring that $C(\rho_0) = \langle U^2 \rangle$. For $L=0$, this yields $\rho_0 \sim (1-\Delta)^{-1/2}$. This is readily verified by the data in Fig.\ref{Fig_spatial}A: As the system approaches the onset of collective motion, the overall strength of the correlations grows, with the region of strong correlations extending to progressively larger scales. 

\begin{figure}[t]
\centering
\includegraphics[width=0.8\linewidth]{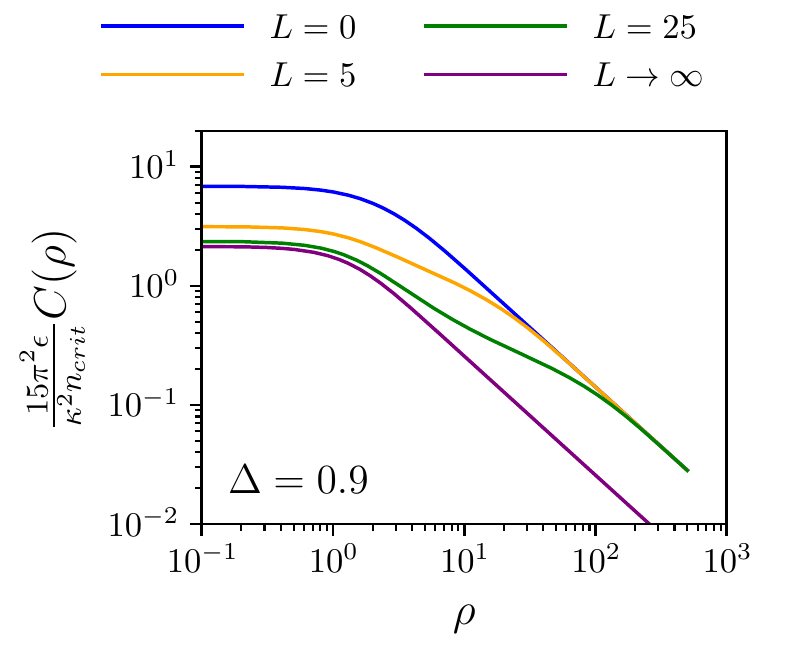}
\caption{The spatial correlation function $C(\rho)$ as a function of the distance $\rho$ for $\Delta=0.9$ and various values of $L$. At sufficiently large distances, $C(\rho)$ recovers the shaker behaviour, while at small distances correlations are suppressed by swimming. Note that the $L\rightarrow\infty$ line, serving as the limit beyond which correlations cannot be suppressed, joins the shaker line at $\rho\rightarrow\infty$.}
\label{Fig_spatial_delta09}
\end{figure}

The effect of swimming on the behaviour of $C(\rho)$ is demonstrated in Figs.\ref{Fig_spatial}B-\ref{Fig_spatial}C. As $L$ increases, the strongly correlated core at moderate separations shrinks, indicating that the steady growth of orientational correlations is reduced by the mixing introduced by swimming. The overall strength of correlations inside the core also decreases with $L$, reflecting the reduction of the fluid velocity variance by swimming. At large distances, $C(\rho)$ recovers the behaviour seen in shakers, with the crossover distance given by $\rho_1 \sim L (1-\Delta)^{-1/2}$, as can be deduced from the exponential in Eq.\eqref{spatialcorrelationsanalytic}. This behaviour is further demonstrated in Fig.\ref{Fig_spatial_delta09}, where we plot $C(\rho)$ for $\Delta=0.9$ and various values of $L$. In the limit of fast swimming, $L\rightarrow\infty$, the correlation function deviates modestly from the non-interacting case for almost all values of $\Delta$, exhibiting a quick rise and the divergence associated with the onset of collective motion only in a very small vicinity of $\Delta=1$.

The data in Fig.\ref{Fig_spatial} and Eq.\eqref{spatialcorrelationsanalytic} demonstrate that $C(\rho)$ exhibits an algebraic decay for large distances, and a true correlation length can thus not be defined. A phenomenological correlation length $\eta_\mathrm{corr}$ can nevertheless be defined as a distance over which $C(\rho)$ decreases by certain amount, as has been employed in \cite{Gachelin2014, Bardfalvy2019}. Setting $C(\eta_\mathrm{corr}) = \alpha \langle U^2 \rangle$, with $\alpha<1$, we obtain 
\begin{align}
\eta_\mathrm{corr}\sim (1-\Delta)^{-1/2}, \quad \Delta\rightarrow1, 
\end{align}
similar to any other typical distance discussed above.

\subsection{Fluid velocity spectrum}

\begin{figure*}
\centering
\includegraphics[width=1\linewidth]{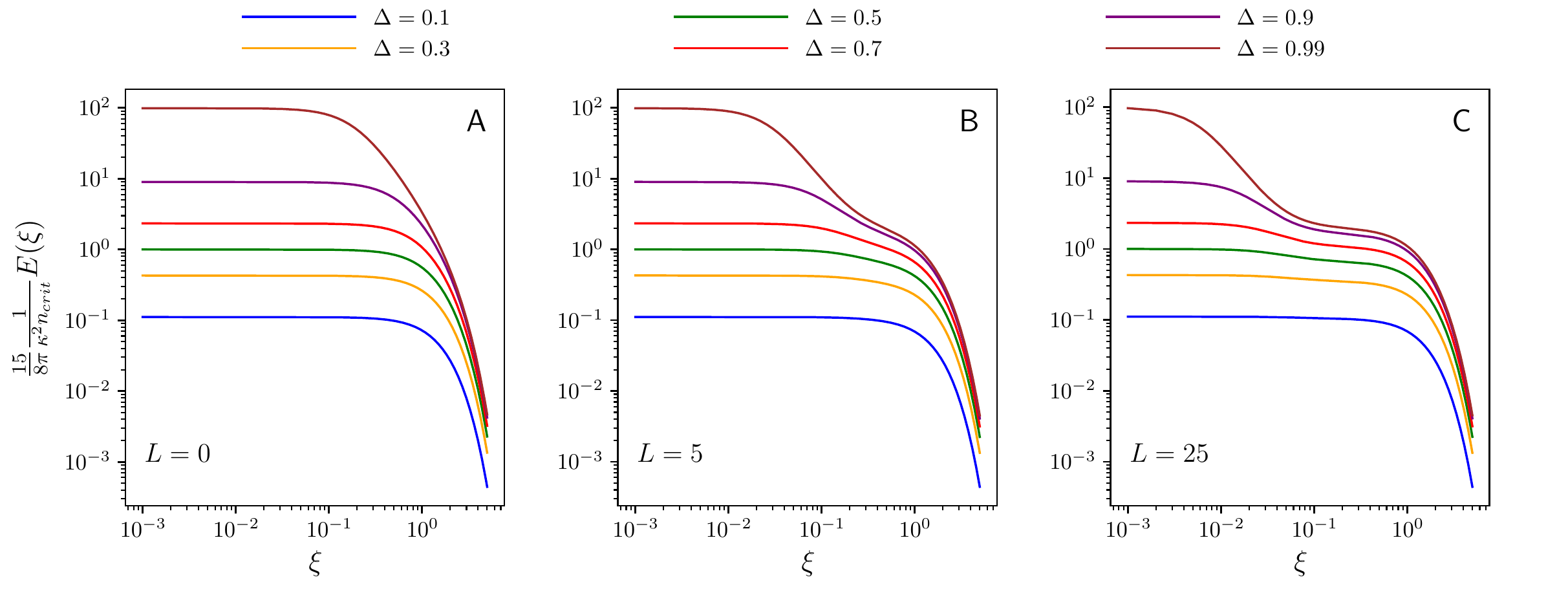}
\caption{The fluid velocity energy spectra $E(\xi)$, Eq.\eqref{energyspectrum}, as a function of the dimensionless wavenumber $\xi$ for various values of $\Delta$. A: $L=0$, B: $L=5$, and C: $L=25$. The legend applies to all panels.}
\label{Fig_Spectra}
\end{figure*}

Next, we discuss the fluid velocity energy spectrum $E(k)$ that is closely related to the spatial correlation function $C(\rho)$. Defined as
\begin{align}
E(k) = 4\pi k^2 \overline{\hat{U}^{\alpha}\left( {\bm k} \right) \hat{U}^{\alpha}\left( -{\bm k} \right)},
\label{eq:spectrum}
\end{align}
this quantity is often used in turbulence research to study the cascade of the kinetic energy \cite{Townsend1980}. Although the kinetic energy is not a useful concept for Stokesian flows, $E(k)$ provides an insight into the relative strength of fluid motion at various scales. The energy spectrum is proportional to the Fourier transform of $C(\rho)$, and, up to a prefactor is given by the integrand of Eq.\eqref{spatialcorrelations}
\begin{align}
&E(\xi) = \frac{8\pi}{15} \kappa^2 n A^2(\xi) \nonumber \\
&\qquad\qquad \times \frac{2-A(\xi)\Delta+\frac{6}{7}L^2 \xi^2}{\left(2-A(\xi)\Delta\right)\left(1-A(\xi)\Delta+\frac{3}{7}L^2 \xi^2\right)},
\label{energyspectrum}
\end{align}
where, again, $\xi=k\epsilon$. This expression is plotted in Fig.\ref{Fig_Spectra} for various values of $\Delta$ and $L$.

First, we observe that $E(\xi)$ has significant energy content at all large scales, $\xi<1$, that quickly decays to zero at the organism-size scales, $\xi\sim1$, due to the regularising factor $A(\xi)$. This is not caused by some form of energy cascade, but is due to the nature of the dipolar field created by the microswimmers. Indeed, the dipolar velocity field decays in space as $r^{-2}$, while its Fourier transform scales as $k^{-1}$. Together with the definition of $E(k)$, Eq.\eqref{eq:spectrum}, this implies that $E(k)\sim k^0$ even for a single microswimmer, i.e. the dipolar field has a constant energy content at every scale.  

In the presence of interactions, the energy spectrum of shakers ($L=0$) preserves the overall structure described above, while its absolute value increases with $\Delta$ and, eventually, diverges at $\Delta=1$. For swimmers, the increase in the energy content is mostly confined to large scales, while in the limit of fast swimming (not shown), the rise in the energy content on the approach to the onset of collective motion is confined to the largest scales available ($k\rightarrow0$) and starts to be visible only in a very close vicinity of $\Delta=1$.

\subsection{Temporal correlations}
\label{section:results:temporal}

\begin{figure*}
\centering
\includegraphics[width=1.0\linewidth]{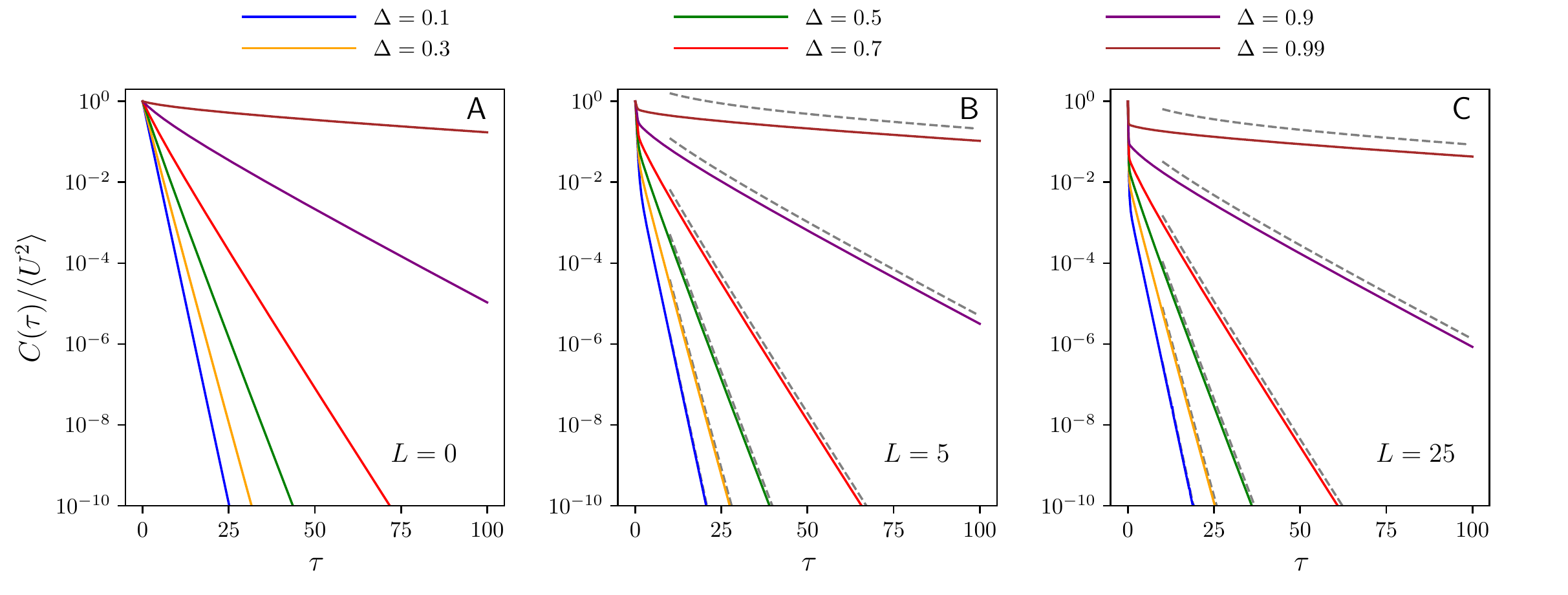}
\caption{The temporal correlation function $C(\tau)$ as a function of the dimensionless time $\tau$ for various values of $\Delta$. A: $L=0$, B: $L=5$, and C: $L=25$. The solid lines are calculated by numerically evaluating Eqs.\eqref{ResultsC0} and \eqref{C1final}, while the dashed lines in B and C are the analytic approximation of the asymptotic behaviour for $\tau\rightarrow\infty$, Eq.\eqref{C1asymptotics}.  The legend applies to all panels.}
\label{Fig_Temporal}
\end{figure*}

The temporal correlation function $C(\tau)=C_0(\rho=0,\tau)+C_1(\rho=0,\tau)$ is given by Eqs.\eqref{ResultsC0} and \eqref{C1final}. The corresponding expressions do not simplify significantly in the limit $\rho=0$, and we do not repeat them here. In Fig.\ref{Fig_Temporal} we plot $C(\tau)$ normalised by its value at $\tau=0$, which is given by the fluid velocity variance $\langle U^2\rangle$. As with the other quantities discussed above, the temporal correlation function exhibits a progressively slower decay as $\Delta$ approaches the onset of collective motion, eventually diverging at $\Delta=1$.
For swimmers, this is offset by a decay of $C(\tau)$ at short times that becomes more pronounced as $L$ increases. For very large swimming speeds, the temporal correlations differ only marginally from the non-interacting case for most values of $\Delta$, eventually exhibiting a rapid increase and divergence in a very small vicinity of $\Delta=1$.

To understand the behaviour of $C(\tau)$ at long times, we analyse its individual contributions. The integral in the non-interacting part, $C_0(T)$, can be explicitly evaluated giving
\begin{align}
& C_0(\tau) = \frac{n \kappa^2 }{\pi \epsilon} \frac{e^{-\tau}}{8\alpha^4 (4+\alpha^2)^2} 
\Bigg[ \nonumber \\
& \qquad\qquad\qquad 4(24+8\alpha^2+\alpha^4) \mathbb{E}\left( -\frac{\alpha^2}{4}\right) \nonumber \\
& \qquad\qquad\qquad-(4+\alpha^2)(24+5\alpha^2) \mathbb{K}\left( -\frac{\alpha^2}{4}\right)
\Bigg],
\label{}
\end{align}
where $\alpha=L\tau$, and $\mathbb{K}(x)$ and $\mathbb{E}(x)$ are the complete elliptic integrals of the first and second order, respectively. In the limits of small and large $\alpha$ this equation predicts
\begin{align}
C_0(\tau) \sim \frac{n \kappa^2 }{\pi \epsilon} e^{-\tau} \times
\begin{cases}
\frac{21\pi}{2048},\quad &L\tau\rightarrow0, \\ \\
\frac{1}{4 (L\tau)^3},\quad &L\tau\rightarrow\infty.
\end{cases}
\label{eq:C0taunoninteracting}
\end{align}
 At short times, tumbling is the leading source of decorrelation, while at large $\tau$ the non-interacting temporal correlation function $C_0$ decays as $\tau^{-3}e^{-\tau}$, as reported previously \cite{Belan2019,Bardfalvy2019}. The crossover time is set by $\alpha=L\tau=v_s t/\epsilon\sim1$, and corresponds to the time interval needed for a microswimmer to swim its own size.

To understand the large-$\tau$ asymptotic behaviour of $C_1(\tau)$, we observe that 
\begin{align}
e^{-\tau} \int_0^\infty d\xi A(\xi)^2 \left\{ 
\begin{array}{c}
\sin{\gamma \tau \xi} \\
\cos{\gamma \tau \xi}
\end{array} \right\} 
\underset{\tau\rightarrow\infty}{\sim}
e^{-\tau}  \left\{ 
\begin{array}{c}
\tau^{-1}\\
\tau^{-5}
\end{array} \right\},
\end{align}
where $\gamma$ is a real constant. This result implies that a trigonometric function in the integrand of Eq.\eqref{C1final} generates a contribution to $C_1(\tau)$ that decays on the same timescale as the non-interacting part $C_0(\tau)$, and does not contribute to the slow decay in Fig.\ref{Fig_Temporal}. In turn, this restricts the integration domain to $\xi \in [0,\xi_*]$, with
\begin{align}
\xi_* = \sqrt{\frac{7}{12}}\frac{\Delta}{L},
\end{align}
which ensures that the arguments of the hyperbolic functions in Eq.\eqref{C1final} are real. Introducing $\zeta = \xi/\xi_*$, $C_1(\tau)$ can be approximated as
\begin{align}
&C_1(\tau)
\underset{\tau\rightarrow\infty}{\sim}
  \frac{n \kappa^2 }{15 \pi^2 \epsilon} e^{-\tau\left(1-\frac{1}{2}\Delta\right)} \xi_* \int_{0}^1 d\zeta  \frac{1}{1-\Delta + \frac{1}{4}\Delta^2\zeta^2} \nonumber \\
& \qquad \times \Bigg\{
\frac{2-\Delta +\frac{1}{2}\Delta^2\zeta^2}{2-\Delta}\cosh{\left(\frac{1}{2}\Delta \tau  \sqrt{1-\zeta^2}\right)} \nonumber \\
&\qquad\qquad\qquad + \frac{1}{\sqrt{1-\zeta^2}} \sinh{\left(\frac{1}{2}\Delta \tau  \sqrt{1-\zeta^2}\right)} 
\Bigg\},
%\label{a}
\end{align} 
where we used $A(\xi<\xi_*)\sim 1$ for not-too-small values of $L$. In the limit of large $\tau$, this can be further approximated by
\begin{align}
&C_1(\tau)
\underset{\tau\rightarrow\infty}{\sim}
  \frac{n \kappa^2 }{15 \pi^2 \epsilon} e^{-\tau\left(1-\Delta\right)} \frac{\xi_*}{1-\Delta} \int_{0}^1 d\zeta  \,
e^{-\frac{1}{4}\tau \Delta \zeta^2} \nonumber \\
& =  \frac{n \kappa^2 }{15 \pi^2 \epsilon}  \sqrt{\frac{7\pi}{12}\frac{\Delta}{\tau}} \frac{1}{L(1-\Delta)} e^{-\tau\left(1-\Delta\right)} \mathrm{erf}\left( \frac{1}{2}\tau\Delta\right),
\label{C1asymptotics}
\end{align} 
where $\mathrm{erf}(x)$ denotes the error function. Predictions of Eq.\eqref{C1asymptotics} are plotted in Fig.\ref{Fig_Temporal}B and C as dashed lines. We find a good agreement between its prediction and the true decay of $C(\tau)$ as $\tau\rightarrow\infty$. 

To extract the typical timescale $\tau_{corr}$ of the fluid velocity fluctuations on the approach to collective motion, we combine Eqs.\eqref{varianceasymptotic} and \eqref{C1asymptotics} to obtain
\begin{align}
\frac{C(\tau)}{\langle U^2\rangle}
\underset{\tau\rightarrow\infty}{\sim}
\frac{e^{-\tau\left(1-\Delta\right)}}{\sqrt{\tau \left(1-\Delta\right)}}, \quad \Delta \rightarrow 1,
\end{align}
which implies
\begin{align}
\tau_{corr} \sim (1-\Delta)^{-1}.
\end{align}

\subsection{Enhanced diffusivity}

As the final observable, we consider here the enhanced diffusivity of a passive tracer particle embedded in a suspension of motile microorganisms. The tracer is assumed to be neutrally buoyant and move due to advection by the velocity fields created by the microswimmers. Brownian diffusion of the tracer is significantly weaker than its enhanced counterpart, and is neglected for simplicity. This problem has been extensively studied both experimentally \cite{Wu2000,Kim2004,Leptos2009,Kurtuldu2011,Mino2011,Mino2013,Jepson2013,Patteson2016} and theoretically \cite{Underhill2008,Dunkel2010,Ishikawa2010,Childress2010,Childress2011,Pushkin2013,Pushkin2013jfm,Morozov2014,Kasyap2014,Thiffeault2015,Burkholder2017} in the dilute regime, where $\Delta\ll 1$, and for arbitrary densities of shakers \cite{Stenhammar2017}. Here, we consider the case of arbitrary density $\Delta < 1$ and $L$.

The position of the tracer ${\bm a}(T)$ obeys the following equation of motion
\begin{align}
\dot{\bm a}(T) = {\bm U}({\bm a}(T),T),
\label{tracer_position}
\end{align}
which implies that the tracer is point-like and follows the velocity of the fluid at its position. The long-time behaviour of such a tracer is diffusive \cite{Wu2000,Leptos2009,Thiffeault2015}, and the associated diffusion coefficient can be extracted in the usual way
\begin{align}
D = \lim_{T\rightarrow\infty} \frac{1}{6T} \overline{{\bm a}(T)\cdot {\bm a}(T)}.
\end{align}
Here, the bar denotes the average over the history of tumble events, and has the same meaning as in Eq.\eqref{CRTgeneral}. Solving formally Eq.\eqref{tracer_position}, ${\bm a}(T)={\bm a}(0) + \int_0^T dt' {\bm U}({\bm a}(t'),t')$, the diffusion coefficient can be written as \cite{Kubo1966}
%\begin{align}
%&D = \frac{1}{3} \int_{0}^\infty dT\, \overline{{\bm U}({\bm a}(t+T),T) \cdot {\bm U}({\bm a}(t),t)}\nonumber \\
%&\qquad\qquad \approx \frac{1}{3} \int_{0}^\infty dT\, C(T).
%\end{align}
\begin{align}
&D = \frac{1}{3} \lim_{t\rightarrow\infty}\int_{0}^\infty dT\, \overline{{\bm U}({\bm a}(t+T),t+T) \cdot {\bm U}({\bm a}(t),t)}.
\end{align}
Here, $t$ is sufficiently large so that any influence of the initial conditions has died away. To proceed, we observe that ${\bm U}({\bm a}(t+T),t+T)$ can be iteratively calculated by substituting the formal solution for ${\bm a}(T)$ into its spatial argument, i.e.
\begin{align}
&{\bm U}({\bm a}(t+T),t+T) = {\bm U}({\bm a}(t),t+T)\nonumber \\
& +\nabla  {\bm U}({\bm a}(t),t+T) \cdot \int_{t}^{t+T} dt' {\bm U}({\bm a}(t'),t') + \cdots.
\label{DUapprox}
\end{align}
As was argued by Pushkin and Yeomans \cite{Pushkin2013}, for very dilute suspensions velocity gradients over the typical distance travelled by the tracer particle during the microswimmer runtime are small compared to the velocity of the fluid at any of these positions, and can be neglected. Therefore, we can approximate the diffusion coefficient as
\begin{align}
& D \approx \frac{1}{3} \int_{0}^\infty dT\, \overline{{\bm U}({\bm a}(t),t+T) \cdot {\bm U}({\bm a}(t),t)}\nonumber \\
&\qquad\qquad = \frac{1}{3} \int_{0}^\infty dT\, C(T).
\label{DvsCT}
\end{align}
As we have seen in Section \ref{section:results:temporal}, as $\Delta$ increases, the correlation time increases from $\lambda^{-1}$ (corresponding to $\tau_{corr}=1$) in the very dilute regime to progressively larger values, eventually diverging as $\Delta\rightarrow 1$, implying that the second, etc. terms in Eq.\eqref{DUapprox} grow rapidly in this limit. However, the fluid velocity variance, which sets the magnitude of the leading term in Eq.\eqref{DUapprox} also diverges as $\Delta\rightarrow 1$. Further work is required to assess the validity of the approximation above for all values of $\Delta$. Here, we proceed by using Eq.\eqref{DvsCT} with the potential caveat that it might not be accurate in the vicinity of $\Delta=1$.

The integral in Eq.\eqref{DvsCT} can be evaluated explicitly, leading to $D = D_0 + D_1$, where the non-interacting and interacting contributions are given by
\begin{align}
D_0 = \frac{\kappa^2 n}{45 \pi^2 \lambda \epsilon} \int_{0}^\infty d\xi A^2(\xi) \psi\left( \xi L\right),
%\frac{2x^3 + 3x - 3\left( 1+x^2\right)\arctan{x}}{x^5},
\end{align}
and
\begin{align}
&D_1 = \frac{\kappa^2 n \Delta}{45 \pi^2 \lambda \epsilon} \int_{0}^\infty d\xi A^3(\xi) \nonumber\\
&\qquad\qquad \times \frac{ 2-A(\xi)\Delta+\frac{6}{7}L^2 \xi^2 }{\left( 1+\frac{3}{7} L^2\xi^2\right)\left( 1-A(\xi)\Delta+\frac{3}{7}L^2 \xi^2\right)^2},
\end{align}
respectively, and $\psi(x)$ is defined in Eq.\eqref{psi}. At this point, we would like to comment on the shaker limit of these expressions, when they should reduce to the ones obtained by Stenhammar \emph{et al.} \cite{Stenhammar2017}. Instead, we observe that the expression for $D_1$ reported there erroneously contained $A^2(\xi)$ instead of $A^3(\xi)$ under the integral. We note, however, that since $A(\xi)$ is a regularised representation of a step function, this has almost no bearing on the numerical evaluation of  $D_1$ presented in \cite{Stenhammar2017}.

The integral in the non-interacting part $D_0$ cannot be represented in terms of special functions, but its limiting behaviour can readily be obtained. Combining the asymptotic results for $L=0$ and $L\rightarrow\infty$, results in the following approximation
\begin{align}
D_0 \approx \frac{\kappa^2 n}{\lambda \epsilon} \frac{7}{2048 + 336 \pi L}.
\label{D0approx}
\end{align}

To derive an approximate expression for $D_1$, we set $A(\xi)\approx 1$ under the integral sign, to obtain
\begin{align}
D_1 \approx \frac{\kappa^2 n}{90 \pi \lambda \epsilon \sqrt{1+\frac{12}{7}L^2}} \left\{ \frac{2-\Delta}{\left( 1-\Delta\right)^{3/2}} -2\right\}.
\label{D1approx}
\end{align}

\begin{figure*}
\centering
\includegraphics[width=0.9\linewidth]{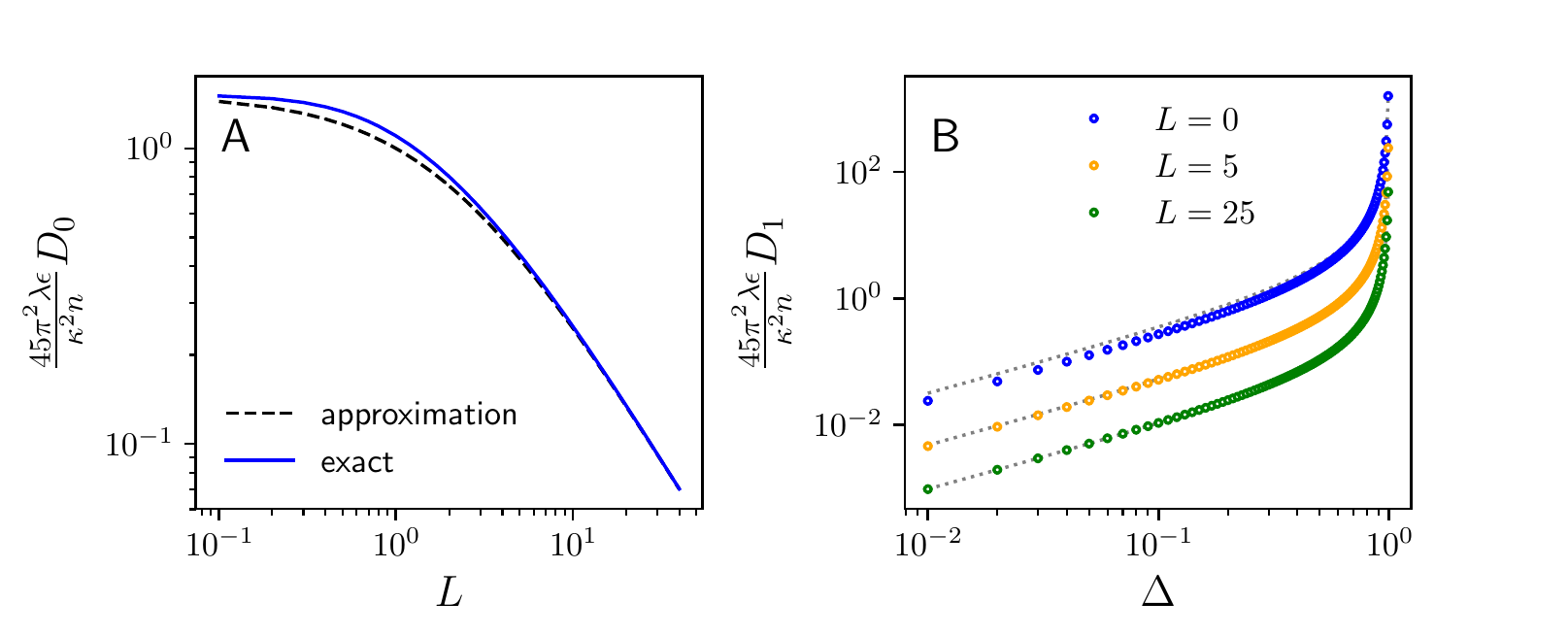}
\caption{A) The non-interacting part of the diffusivity, $D_0$. The dashed line is the approximation, Eq.\eqref{D0approx}, developed in the text. B) The interacting part of the diffusivity, $D_1$. The dotted lines correspond to the approximate expression, Eq.\eqref{D1approx}, evaluated at the corresponding value of $L$.}
\label{Fig_Diffusivity}
\end{figure*}

In Fig.\ref{Fig_Diffusivity} we compare the numerical evaluation of $D_0$ and $D_1$ against Eq.\eqref{D0approx} and \eqref{D1approx}. We observe that while the uniform approximation Eq.\eqref{D0approx} does not work well for small but finite values of $L\sim1$, all other values of $L$ are well-represented by the approximation. The interacting part of the diffusivity is well-approximated by Eq.\eqref{D1approx}.

Finally, we remark that Eq.\eqref{D1approx} predicts that
\begin{align}
D_1 \sim (1-\Delta)^{-3/2}, \quad \Delta\rightarrow1,
\end{align}
even though this prediction should be treated with caution, as discussed above.

\section{Discussion and Conclusion}
\label{section:discussion}

In this work, we have presented a kinetic theory for dilute suspensions of pusher-like microswimmers interacting via long-ranged dipolar fields. {\alex We have overcome a significant technical difficulty in including particle self-propulsion into a theory that goes beyond the mean-field assumption and explicitly accounts for correlations between microswimmers. This difficulty has limited previous theoretical work on this problem to either the case of shaker microswimmers \cite{Stenhammar2017} or the case of swimming being subdominant compared to the translational thermal diffusion \cite{Qian2017}. The only theory to date that has accounted for arbitrary swimming speeds was developed by Nambiar \emph{et al.} \cite{Nambiar2019a}, who analytically considered pair-wise correlations between microswimmers, \emph{i.e.}, their results are  $O(\Delta^2)$ accurate. To deal with the problem posed by the self-propulsion term in their equations, Nambiar \emph{et al.} \cite{Nambiar2019a} developed a perturbation theory in terms of the swimmer slenderness (aspect ratio), which is a reasonable approximation for long and slender bacteria. 
%However, it is \emph{a priori} unclear whether the terms in such an expansion keep their order for higher values of $\Delta$, and it would be challenging to use this method to analyse the behaviour close to the onset of collective motion, i.e., as $\Delta\to 1$. 
In contrast, the method developed in this work allows us to make explicit predictions for various experimentally relevant observables for any strength of self-propulsion and any density of microswimmers up to the onset of collective motion. 
All of its parameters can be independently measured or inferred from experiments, and its predictions can be directly compared against experimental data. }

The results of our theory, presented in Section \ref{section:results}, reveal that all observables considered deviate from their mean-field values, which can be recovered from our results by setting $\Delta=0$, indicating that the mean-field theory is incorrect at any density below the onset of collective motion. We have also uncovered the following interplay between the strength of correlations between microswimmers and their self-propulsion speed. For all observables considered, the strongest correlations are exhibited by suspensions of shakers, $L=0$. This can be readily seen by observing that, in the absence of self-propulsion, the microswimmer positions only change due to their mutual advection. 
In dilute suspension, displacements thus accumulated over one correlation time are small compared to the interparticle distances, and, to first approximation, shaker suspensions perform orientational dynamics only. In turn, this implies that they spend maximum amount of time possible adjusting to the orientational fields created by other microswimmers. In contrast, motile microswimmers are aligning in a local velocity field that constantly changes due to their self-propulsion, implying weaker correlations in such suspensions. This effect becomes stronger as $L$ increases.

The degree to which correlations are suppressed by self-propulsion depends on the nature of the observable. Spatial-like observables (the fluid velocity variance, the energy spectrum, and the spatial correlation function) are significantly reduced as $L$ increases, but do not reach their mean-field values even in the limit $L\rightarrow\infty$. For instance, as can be seen from Fig.\ref{Fig_Variance}, the fluid velocity variance is significantly larger than its mean-field value at any density $\Delta$, even in the limit of fast swimming. In a similar fashion, as $L\rightarrow\infty$, the spatial correlation function in Fig.\ref{Fig_spatial_delta09} does not reduce to its mean-field behaviour, which is given by the $\Delta\rightarrow0$ limit in Fig.\ref{Fig_spatial}. Instead, it recovers the strongly correlated shaker-like behaviour at sufficiently large distances. This can be understood by employing the same argument as above. For any value of $L$, there are such separations $\rho$ that the typical distance travelled by a microswimmer during one correlation time of the suspension is small compared to $\rho$. For such separations, the difference between swimmers and shakers vanishes and $C(\rho)$ recovers its shaker-like behaviour. 

On the other hand, temporal-like observables (the temporal correlation function and the enhanced diffusivity of tracer particles) are almost completely suppressed as $L\rightarrow\infty$ for $\Delta<1$, though they still diverge in the limit of $\Delta\rightarrow1$. This behaviour mirrors the dependence of their mean-field values on $L$, which vanish in the limit of fast swimming below the onset of collective motion. An intuitive argument for this behaviour has been put forward by Dunkel \emph{et al.} \cite{Dunkel2010}, who demonstrated that the total displacement of a tracer by a single motile particle vanishes as the length of a straight path covered by the swimmer diverges. This is fundamentally related to the time-reversibility of Stokesian flows. The presence of correlations between microswimmers breaks this time-reversibility: although the pathway between two states in phase space is still reversible, the probabilities of finding the suspension in those states are a priori different. Strong swimming introduces effective phase space mixing and recovers equal a priori probabilities for the phase space states. Again, this argument only holds for $\Delta<1$ when $L$ is large, yet finite.

{\alex Previous studies have already reported measurements of the spatial \cite{Gachelin2014,Martinez2020} and temporal \cite{Soni2003,Wu2006} correlation functions in dilute bacterial suspensions for a wide range of bacterial concentrations. While these observations qualitatively agree with our predictions and the results of previous simulations \cite{Saintillan2012,Krishnamurthy2015,Bardfalvy2019}, quantitative comparison is problematic as the corresponding values of $\Delta$ in those experiments remain unknown. The enhanced diffusivity, on the other hand, has only been studied in the regime where $D$ scales linearly with the bacterial number density $n$ \cite{Kim2004,Mino2011,Mino2013,Jepson2013,Kasyap2014,Patteson2016}, with the highest density of Kasyap \emph{et al.} \cite{Kasyap2014} being the only exception. Those measurements are well-described by a non-interacting theory \cite{Childress2011,Pushkin2013,Jepson2013,Morozov2014}, i.e., $D_0$ in our analysis, and to test our theory they will need to be extended to higher concentrations. Therefore, to verify our predictions experimentally, it is necessary to measure any of these observables across a wide range of bacterial density while carefully controlling the distance to the threshold of collective motion $\Delta$. Although the latter can, in principle, be calculated from $\Delta = n B\kappa/5\lambda$, it requires the knowledge of the bacterial dipolar strength, tumble rate, and effective aspect ratio, and a precise control of the number density $n$. A significantly easier approach would be to determine $n_{crit}$ experimentally. This can be achieved, for instance, by measuring the apparent shear viscosity of bacterial suspensions at various densities, as was recently done by Martinez \emph{et al.} \cite{Martinez2020}. In sufficiently wide geometries, the ratio of the apparent shear viscosity to the viscosity of the solvent decreases linearly with  $\Delta$ \cite{Saintillan2018,Martinez2020}, and vanishes precisely at the onset of collective motion \cite{Subramanian2009}. Simultaneous measurement of one of the observables discussed above and the apparent shear viscosity would thus allow for a direct comparison with our theory. The remaining parameters, $\epsilon$ and $\lambda$ used to rescale space and time, respectively, and the dimensionless persistence length $L$, should be treated as fitting parameters. They can be fixed, for instance, by fitting the data for very low bacterial number densities, where the normalised correlation functions $C(\rho)/\langle U^2\rangle$ and $C(\tau)/\langle U^2\rangle$ are well-approximated by their non-interacting (i.e. $\Delta=0$) components (see Eqs.\eqref{eq:U2noninteracting}, \eqref{spatialcorrelations}, and \eqref{eq:C0taunoninteracting} and Refs. \cite{Belan2019,Bardfalvy2019}). While the swimming speed $v_s$ and the tumbling rate $\lambda$ can be directly measured by either tracking individual bacteria or by differential dynamic microscopy \cite{Wilson2011}, the hydrodynamic size of an \emph{E.coli} bacterium $\epsilon$ is somewhat open to interpretation, as discussed in Section \ref{section:results}. Therefore, $L$ should be seen as a fitting parameter.}

{\alex Direct verification of our prediction that increasing $L$ suppresses correlations and brings the system closer to the mean-field predictions would require the ability to perform experiments at different values of $L$ at a fixed distance to the threshold $\Delta$. An obvious realisation of this protocol would involve the ability to control the tumbling rate $\lambda$, while keeping the swimming speed and the dipolar strength constant. We are currently not aware of a bacterial strain with such an ability. An interesting alternative would be to employ the recently created \emph{E.coli} mutants that only swim in the presence of light \cite{Walter2007,Arlt2018,Arlt2019}. In such bacteria, the swimming speed can be increased by increasing the light intensity, while the tumbling rate seems to stay constant for upward sweeps in light intensity \cite{Arlt2019}. Performing light intensity sweeps at various bacterial densities would thus trace a set of straight lines in the $\Delta$-$L$ parameter space, since $v_s$ is expected to be proportional to the bacterial dipolar strength and, thus, to $\Delta$. Such data can then be used to of how the transition is approached at various values of $L$.}

\begin{table}[h!]
  \begin{center}
    \caption{Critical exponents}
    \label{table:exponents}
    \begin{tabular}{ll}
      \textbf{Observable} & \qquad\textbf{Scaling law for $\Delta\rightarrow1$} \\
      \hline
      Fluid velocity variance & \qquad $(1-\Delta)^{-1/2}$ \\
      (Pseudo-)correlation length & \qquad$(1-\Delta)^{-1/2}$ \\
      Correlation time & \qquad$(1-\Delta)^{-1}$ \\
      Enhanced diffusivity & \qquad$(1-\Delta)^{-3/2}$ \\
      \hline
    \end{tabular}
  \end{center}
\end{table}

To gain further insight into the nature of the transition to collective motion exhibited by our model, we extracted the scaling behaviour of the observables considered in this work upon the approach to the onset, $\Delta=1$. All of these quantities diverge at $\Delta=1$ and the values of the critical exponents predicted by our theory are summarised in Table \ref{table:exponents}. We want to stress that these exponents rely on the approximation introduced in Section \ref{subsection:approximateDILT}, and while we are confident that it semi-quantitatively captures the spatial and temporal behaviour of the generalised correlation function $C(\rho,\tau)$ for $\Delta<1$, its quality in the close vicinity of $\Delta=1$ is untested. The values presented in Table \ref{table:exponents} should thus be seen as a first step in understanding the nature of this transition. Currently, neither the order of the mean-field transition, nor the influence of strong pre-transitional correlations on the transition are understood, and more work is needed to assess whether collective motion in dilute suspensions of hydrodynamically interacting microswimmers defines a new universality subclass of ``wet" active matter models.

In this work, we have only considered pusher-like microswimmers below the onset of collective motion. Recent simulations suggest \cite{Stenhammar2017,Bardfalvy2019} that suspensions of pullers also exhibit strong correlations, although their effect is opposite to what is observed for pushers. The results presented in this work cannot be used to study this effect, i.e. by replacing $\Delta$ with $-\Delta$ in the relevant expressions. Instead, to extend our theory to pullers, one would have to re-evaluate the long-term behaviour of the approximate double inverse Laplace transform in Section \ref{subsection:approximateDILT} for negative values of $\Delta$.

\begin{acknowledgments}
V.~\v{S}. acknowledges studentship funding from EPSRC Grant No.~EP/L015110/1. C.~N. acknowledges the support of an Aide Investissements d’Avenir du LabEx PALM (ANR-10-LABX-0039-PALM). J.~S. acknowledges funding from the Swedish Research Council (Grant No.~2019-03718).
\end{acknowledgments}

\appendix

%\section{Regularised dipolar field}
%\label{RegularisedDipole}
%
%\begin{align}
%u_{d}^{\alpha}({\bm x}_{i}; {\bm z}_{j}) = \frac{F l}{8\pi\mu} \left[ 3
%\frac{\left( {\bm p}_j\cdot {\bm x}'\right)^2 {\bm x}' + \epsilon^2 \left( {\bm p}_j\cdot {\bm x}'\right) {\bm p}_j }{\left( x'^2 + \epsilon^2 \right)^{5/2}} \right. \nonumber \\
%\left. -\frac{{\bm x}'}{\left( x'^2 + \epsilon^2 \right)^{3/2}} \right]
%\end{align}

\section{Double inverse Laplace transform of an archetypal term}
\label{appendix:Cesare}

Here, we show how to calculate the double inverse Laplace transform of Eq.\eqref{dilt_example}. The derivation of Eq.\eqref{C1final} is similar, though lengthy, and we do not present it here.

We start by observing that the double inverse Laplace transform in Eq.\eqref{dilt_example}, given in terms of two Bromwich integrals \cite{Doetsch1974}, can be written as
\begin{align}\label{app:B-1}
\lim_{t\to\infty}
\int_{\Gamma_1}\frac{ds_1}{2\pi i} \frac{e^{s_1 t}}{\lambda + s_1 + i v_s k (\hat{\bm k}\cdot {\bm p})}J(s_1),
\end{align}
where
\begin{align}\label{app:B-2}
J(s_1)
=
\int_{\Gamma_2}\frac{ds_2}{2\pi i} 
\frac{1}{s_1+s_2}
\frac{e^{s_2 (t+T)}}{{\lambda + s_2 - i v_s k (\hat{\bm k}\cdot {\bm p})}}.
\end{align}
By the definition of the inverse Laplace transform \cite{Doetsch1974}, the contours defining the integrals above have to be chosen such that $\Gamma_2$ passes on the right of $-s_1$ and of $-\lambda +iv_s k(\hat{\bm k}\cdot {\bm p})$, while $\Gamma_1$ should pass on the right of all the poles of $J(s_1)$ and of $-\lambda-i v_s k(\hat{\bm k}\cdot {\bm p})$. Observe that the first condition implies that $\Gamma_2$ should be chosen on the right of $-\Gamma_1$.

Next, we observe that, again from the definition of the inverse Laplace transform, $J(s_1)$ is only defined for $\textrm{Re}(s_1)>0$. To proceed, we follow the method often utilised in plasma physics to describe the Landau damping  \cite{Balescu1975}. We perform the analytic continuation of $J(s_1)$ to purely imaginary values of $s_1$ (recall that the analytic continuation of a complex function defined on an open set is the only function $\hat{J}(s_1)$ that is analytic,  defined on a larger set, and equals $J(s_1)$ on the original set), and replace $J(s_1)$ with $\hat{J}(s_1)$ in Eq. \eqref{app:B-1}. Since $\lambda>0$, the difficulty in performing the analytic continuation of $J(s_1)$ lies in the pole at $s_2=-s_1$ of the integrand from Eq.\eqref{app:B-2}. We, therefore, define
\begin{align}
& \hat{J}(s_1) = \int^* \frac{ds_2}{2\pi i} 
\frac{e^{s_2 (t+T)}}{s_2+\lambda-i v_s k(\hat{\bm k}\cdot {\bm p})} \nonumber \\
& \qquad\quad +\frac{1}{2} \ell(-s_1)
 \frac{e^{s_2 (t+T)}}{s_2+\lambda-i v_s k(\hat{\bm k}\cdot {\bm p})}\Bigg\vert_{s_2=-s_1},
\end{align}
where
\begin{align}
\ell(s_1) = \begin{cases}
0, & \textrm{Re}(s_1)>0, \\
1, & \textrm{Re}(s_1)=0, \\
2, & \textrm{Re}(s_1)<0.
\end{cases}
\end{align}
The meaning of the integral denoted by $\int^* ds_1$ above depends on the sign of $\textrm{Re}(s_1)$: If $\textrm{Re}(s_1)>0$, it is just a standard complex integral over a contour passing on the right of $-s_1$ and of $-\lambda +iv_s k(\hat{\bm k}\cdot {\bm p})$; If  $\textrm{Re}(s_1)=0$, $\int^* ds_1$ stands for a principal value integral; Finally, if $\textrm{Re}(s_1)<0$, $\int^* ds_1$ stands for a standard complex integral over a contour passing on the left of $-s_1$ but on the right of $-\lambda +iv_s k(\hat{\bm k}\cdot {\bm p})$. With the definitions above, it is easy to show that $\hat{J}(s_1)$ is holomorphic in an infinitesimal stripe around $s_1\in \mathbb{R}$. Hence, it is the analytic continuation of $J(s_1)$.

Replacing $J(s_1)$ by $\hat{J}(s_1)$ in Eq.\eqref{app:B-1}, we obtain two terms. The first term, containing $\int^*ds_2$, vanishes for $t\to\infty$, since we are now free to choose the integration contours $\Gamma_1$ and $\Gamma_2$ such that $\textrm{Re}(s_1+s_2)<0$. The other term reads
\begin{align}
& \lim_{t\to\infty} \frac{1}{2}\int_{\Gamma_1}\frac{ds_1}{2\pi i} 
\frac{1}{s_1+\lambda+i v_s k(\hat{\bm k}\cdot {\bm p})} \nonumber \\
&\qquad\qquad\qquad\qquad \times \frac{\ell(-s_1)\,e^{-s_1 T}}{-s_1+\lambda-i v_s k(\hat{\bm k}\cdot {\bm p})}.
\end{align}
Closing the contour at $+\infty$, the only pole contributing to the integral is at $s_1=\lambda-i v_s k(\hat{\bm k}\cdot {\bm p})$, and we obtain
\begin{align}
&\lim_{t\to\infty}
\int_{\Gamma_1}\frac{ds_1}{2\pi i} \frac{e^{s_1 t}}{s_1+\lambda+i v_s k(\hat{\bm k}\cdot {\bm p})}J(s_1) \nonumber \\
&\qquad\qquad\qquad\qquad = \frac{e^{-\lambda T + i v_s k T (\hat{\bm k} \cdot {\bm p})}}{2\lambda}.\nonumber
\end{align}
This completes the proof of the equality in Eq.\eqref{dilt_example}.

\section{Approximating $\psi(z)$}
\label{appendix:psia}

Here, we develop an approximation to $\psi(z)$ from Eq.\eqref{psi}. Our goal is to find a rational function with a pole structure that is similar to the original $\psi(z)$. As discussed in Section \ref{subsection:approximateDILT}, the relevant domain is set by the values of $z$ given by $z = \beta/(1+s/\lambda)$, with $\beta=v_s k/\lambda$ varying from $0$ to $L=v_s/(\lambda \epsilon)=0-25$, and by the real part of $s$ ranging from $-\lambda$ to $0$.

Our starting point are the observations that as $z\rightarrow0$, $\psi(z) \rightarrow 1-3z^2/7$, while for $z\rightarrow\infty$, 
$\psi(z) \rightarrow 0$. Both asymptotic behaviours can be combined into $\psi_a(z)= 7/(7+3z^2)$. Now we show that this is a surprisingly good approximation to $\psi(z)$, both reproducing its global shape and having a similar pole structure.

\begin{figure}
\centering
\includegraphics[width=0.8\linewidth]{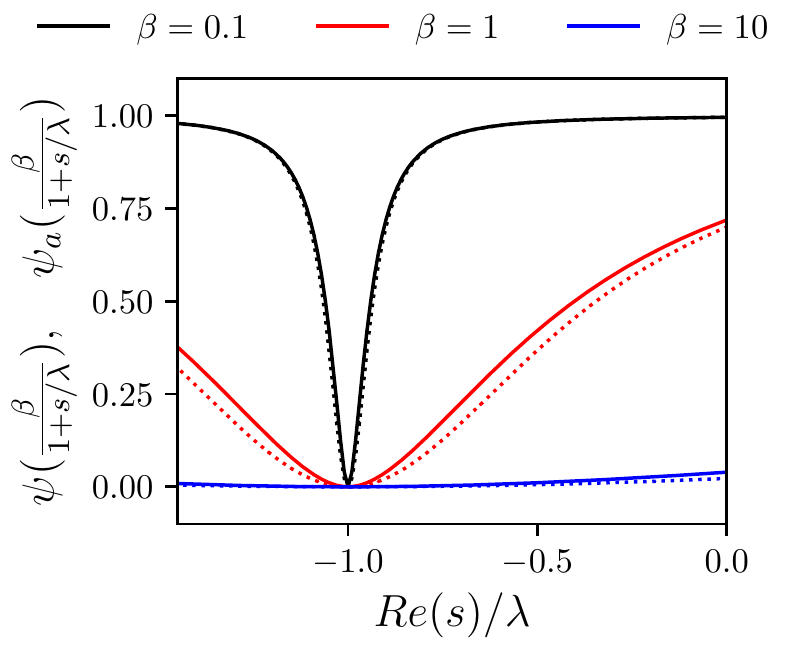}
\caption{Comparison between $\psi(z)$  (solid lines) and $\psi_a(z)$ (dotted lines) for $z = \beta/(1+s/\lambda)$ for real values of $s$ and various values of $\beta$.}
\label{psi_vs_psia}
\end{figure}

In Fig.\ref{psi_vs_psia} we compare $\psi(z)$ and $\psi_a(z)$ for real values of $s$. We observe a good agreement between the two functions for various values of $\beta$. Similar, semi-quantitative, degree of agreement is observed for larger values of $\beta$ and also for complex values $s$. 

To demonstrate that $\psi_a(z)$ also reproduces the pole structure of $\psi(z)$, we consider a typical term from the analysis 
in Section \ref{subsection:hhat}
\begin{align}
\frac{1}{s+\lambda-\lambda\Delta A(k\epsilon)\psi(z)} = \frac{1}{\lambda \Delta A(k\epsilon)}\frac{z}{\omega - z \psi(z)}.
\label{A_original}
\end{align}
We compute its inverse Laplace transform numerically, using the original function $\psi(z)$, and compare the result with the analytic expression, which we obtain by replacing $\psi(z)$ with $\psi_a(z)$ in the expression above. The latter is straightforward: factorising $\omega \left(7+3z^2\right) - 7 z = 3\omega(z-z_{+})(z-z_{-})$, where $z_{\pm}$ are given in Eq.\eqref{z012}, we obtain
\begin{align}
&\frac{1}{\lambda \Delta A(k\epsilon)}\frac{z}{\omega - z \psi_a(z)} \nonumber \\
&\qquad\qquad = \frac{1}{7}\frac{1}{s+\lambda}\frac{7(s+\lambda)^2 + 3(v_s k)^2}{\left(s+\lambda - \frac{v_s k}{z_{+}}\right)\left(s+\lambda - \frac{v_s k}{z_{-}}\right)}.
\end{align}
Performing the inverse Laplace transform of this expression and introducing the dimensionless time $\tau = \lambda t$ yields
\begin{align}
&e^{-\tau} \Bigg[ 1 +\frac{2}{\sqrt{1-\frac{12\beta^2}{7\Delta^2}}} \exp{\left(\frac{\tau \Delta }{2}\right)} \nonumber \\
& \qquad\qquad\qquad\qquad \times \sinh{\left(\frac{\tau \Delta}{2}\sqrt{1-\frac{12\beta^2}{7\Delta^2}}\right)}\Bigg].
\label{A_test_analytic}
\end{align}
Since both $A(k\epsilon)$ and $\Delta$ take values between $0$ and $1$, for the purpose of comparing to its numerical counterpart, we set $A(k\epsilon)=1$ in the expression above, without loss of generality.

\begin{figure*}
\centering
\includegraphics[width=1.0\linewidth]{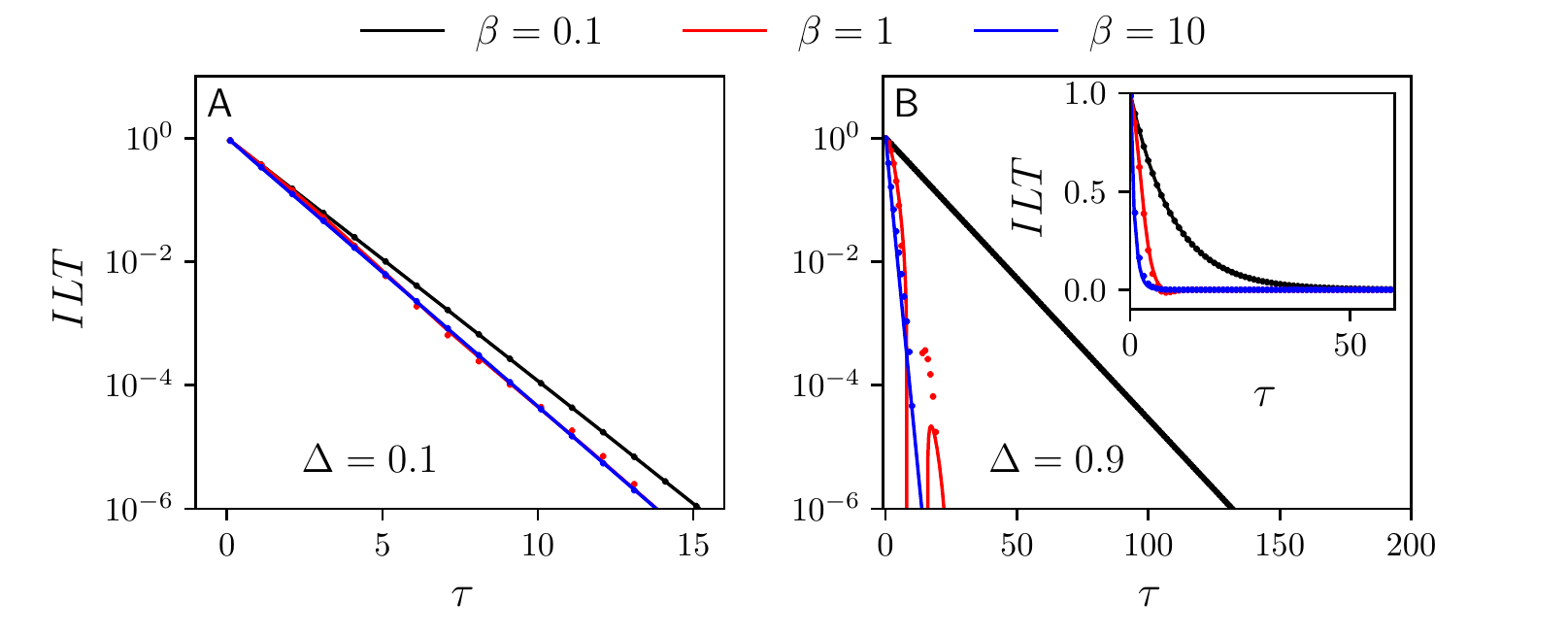}
\caption{Numerical inverse Laplace transform (ILT) of Eq.\eqref{A_test_numerics} (circles) and the analytical approximation, Eq.\eqref{A_test_analytic} (solid lines) as functions of time. A: $\Delta = 0.1$. B: $\Delta = 0.9$. (inset) the same data on the linear-linear scale.}
\label{ILTcombined}
\end{figure*}

The inverse Laplace transform of the original function Eq.\eqref{A_original} written in terms of the same parameters is given by the Bromwich integral
\begin{align}
\frac{1}{2\pi i} \int_{\gamma-i \infty}^{\gamma+i \infty} d{\tilde s} \frac{e^{\tilde s \tau}}{{\tilde s}+1-\Delta \psi(\beta/(\tilde s+1))},
\label{A_test_numerics}
\end{align}
where $\gamma$ is a real number, chosen to be greater than the real part of any singularity of the integrand \cite{Doetsch1974}. We perform this integral numerically, using the Gaver–Wynn–Rho algorithm as presented by Valko and Abate \cite{Valko2005}. Valko and Abate provide an explicit Mathematica function GWR \cite{math1}, which we use here. A Mathematica notebook with the details of this calculation can be found here \cite{SI}. 

In Fig.\ref{ILTcombined}A we compare Eq.\eqref{A_test_analytic} against the numerical Laplace transform of Eq.\eqref{A_test_numerics} for $\Delta=0.1$ and $\beta=0.1$, $1$, and $10$. We observe a very good agreement, which is not surprising: At small microswimmer densities, the hydrodynamic interactions between particles affect their dynamics only weakly, and correlations decay as $e^{-\tau}$. This regime does not test the quality of our approximation. A more stringent test is provided, on the other hand, in Fig.\ref{ILTcombined}B, were we compare the two Laplace transforms for $\Delta=0.9$. For $\beta<1$, we observe a very good agreement even at such high values of $\Delta$ (close to the mean-field transition). This is the most interesting regime, corresponding to large-scale motion in the suspension, and it is encouraging that our approximation shows quantitative agreement with the numerical data. Note that the black line and the black circles, corresponding to $\beta=0.1$, do not follow $e^{-\tau}$, i.e. our approximation is capable of capturing a non-trivial decay rate. At higher values of $\beta$, corresponding to scales comparable to individual microswimmers, the agreement is semi-quantative, but the overall decay is again close to the tumbling-dominated decay $e^{-\tau}$.

In Appendix \ref{appendix:dilt}, we assess the quality of our approximation, when used in Eq.\eqref{C1laplace}, which is its ultimate purpose.

\section{Double inverse Laplace transform}
\label{appendix:dilt}

In Section \ref{subsection:approximateDILT}, we performed the double inverse Laplace transform in Eq.\eqref{C1laplace} analytically by replacing $\psi(z)$ with $\psi_a(z)$, which led to Eq.\eqref{C1final}. Here, we assess the quality of that approximation by performing the double inverse Laplace transform in Eq.\eqref{C1laplace} numerically. The relevant part of Eq.\eqref{C1laplace} reads
\begin{align}
&2{\mathcal L}^{-1}_{\tilde{s}_1,\tilde{t}} {\mathcal L}^{-1}_{\tilde{s}_2,\tilde{t}+\tau}
 \frac{1}{1 + \tilde{s}_1}\frac{1}{1 + \tilde{s}_2}\frac{1}{\tilde{s}_1 + \tilde{s}_2} \frac{\tilde{z}_1 \psi(\tilde{z}_1) + \tilde{z}_2 \psi(\tilde{z}_2)}{\tilde{z}_1+\tilde{z}_2}
 \nonumber \\
&\qquad \times\Bigg[ \frac{\tilde{z}_1 \psi(\tilde{z}_1)}{\tilde{\omega} - \tilde{z}_1 \psi(\tilde{z}_1)} + \frac{\tilde{z}_2 \psi(\tilde{z}_2)}{\tilde{\omega} - \tilde{z}_2 \psi(\tilde{z}_2)} \nonumber \\
&\qquad\qquad 
+ \frac{\tilde{z}_1 \tilde{z}_2 \psi(\tilde{z}_1)\psi(\tilde{z}_2)}{\left(\tilde{\omega} - \tilde{z}_1 \psi(\tilde{z}_1)\right)\left(\tilde{\omega} - \tilde{z}_2 \psi(\tilde{z}_2)\right)} \Bigg],
\label{DILTeqnum}
\end{align}
where, in anticipation of performing numerical calculations, we introduced the dimensionless times $\tau = \lambda T$ and ${\tilde t} = \lambda t$, Laplace frequencies ${\tilde s}_{1,2} = s_{1,2}/\lambda$, ${\tilde z}_{1,2} = \beta/(1+{\tilde s}_{1,2})$, and $\tilde{\omega} = \beta/\Delta$, where we absorbed $A(k\epsilon)$ into $\Delta$, as in Appendix \ref{appendix:psia}. In what follows, we set $\tilde t=20$ to imitate the limit $\tilde{t}\rightarrow\infty$. The calculations are performed in Mathematics using the combined Fixed-Talbot and Gaver–Wynn–Rho algorithm described by Valko and Abate
\cite{Valko2005}. A Mathematica notebook with the details of this calculation can be found here \cite{SI}. The results are compared to the relevant part of Eq.\eqref{C1final}, recast in the same dimensionless variables
\begin{align}
&e^{-\tau} \Bigg[ -\cos{\left(\sqrt{\frac{3}{7}}\beta\tau\right)} + \frac{e^{\frac{1}{2} \Delta \tau}}{1-\Delta + \frac{3}{7}\beta^2} \Bigg\{ \nonumber \\
&  \qquad\qquad \frac{2-\Delta + \frac{6}{7}\beta^2}{2-\Delta}\cosh{\left(\frac{1}{2}\Delta \tau  \sqrt{1-\frac{12 \beta^2}{7 \Delta^2}}\right)} \nonumber \\
&\qquad\qquad\qquad +\frac{\sinh{\left(\frac{1}{2} \Delta \tau  \sqrt{1-\frac{12 \beta^2}{7 \Delta^2}}\right)}}{\sqrt{1-\frac{12 \beta^2}{7 \Delta^2}}}
\Bigg\}
\Bigg].
\label{DILTeqanal}
\end{align} 

\begin{figure*}
\centering
\includegraphics[width=1.0\linewidth]{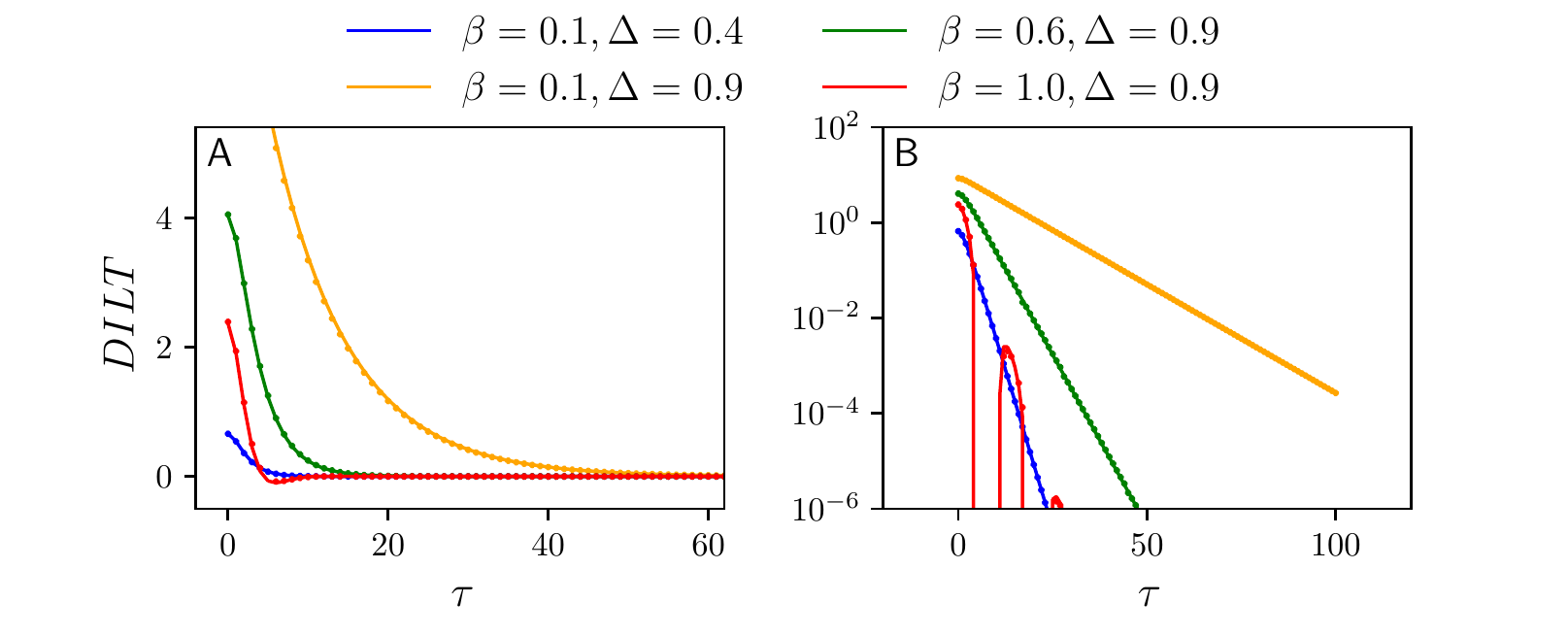}
\caption{Numerical double inverse Laplace transform (DILT) of Eq.\eqref{DILTeqnum} (circles) and its analytical counterpart, Eq.\eqref{DILTeqanal} (solid lines), as functions of time. A: linear- and B: log-linear scales.  The legend applies to both panels.}
\label{DILTcombined}
\end{figure*}

The results of the numerical double inverse Laplace transform and its analytical counterpart are shown in Fig.\ref{DILTcombined}. As in Appendix \ref{appendix:psia}, we focus on high values of $\Delta$, which provide the most stringent test of our results. For $\beta\le1$, the analytic approximation agrees quite well with the numerical data, capturing not only the decay rate, but also the oscillatory behaviour, as can be seen from the $\beta=1$ case. These calculations required a very high number of terms, $O(100)$, in the combined Fixed-Talbot and Gaver–Wynn–Rho algorithm \cite{Valko2005}. For $\beta > 1$, we were unable to obtain converged results for the numerical Laplace transform for any viable number of terms in the numerical algorithm. Nevertheless, the results of Appendix \ref{appendix:psia}, and the degree of agreement exhibited in Fig.\ref{DILTcombined} for the physically most relevant case of $\beta<1$ make us confident that 
Eq.\eqref{C1final} faithfully reproduces the long-time behaviour of Eq.\eqref{C1laplace}.

%\section{Acknowledgement}
%
%Further research outputs generated for this project can be found at http://dx.doi.org/xxxxxx.
%

\bibliography{Refs.bib}

\end{document}